\documentclass[twocolumn]{aastex631}

\usepackage{multirow}

\newcommand{\rtwo}{\ensuremath{R_{21}}}
\usepackage{amsmath}
 \usepackage{pifont}
\usepackage{CJKutf8}

\newcommand{\Princeton}{\affiliation{Department of Astrophysical Sciences, Princeton University, 4 Ivy Lane, Princeton, NJ 08544, USA}}

\newcommand{\OSU}{\affiliation{Department of Astronomy, The Ohio State University, 140 West 18th Avenue, Columbus, OH 43210, USA}}

\newcommand{\CCAPP}{\affiliation{Center for Cosmology and Astroparticle Physics (CCAPP), 191 West Woodruff Avenue, Columbus, OH 43210, USA}}

\newcommand{\Alberta}{\affiliation{Department of Physics, University of Alberta, Edmonton, AB T6G 2E1, Canada}}

\newcommand{\Bonn}{\affiliation{Argelander-Institut f\"ur Astronomie, Universit\"at Bonn, Auf dem H\"ugel 71, 53121 Bonn, Germany}}

\newcommand{\CfA}{\affiliation{Center for Astrophysics $\mid$ Harvard \& Smithsonian, 60 Garden Street, Cambridge, MA 02138, USA}}

\newcommand{\Connecticut}{\affiliation{Department of Physics, University of Connecticut, 196A Auditorium Road, Storrs, CT 06269, USA}}

\newcommand{\ESO}{\affiliation{European Southern Observatory, Karl-Schwarzschild Stra{\ss}e 2, D-85748 Garching bei M\"{u}nchen, Germany}}

\newcommand{\MPE}{\affiliation{Max-Planck-Institut f\"{u}r extraterrestrische Physik, Giessenbachstra{\ss}e 1, D-85748 Garching, Germany}}

\newcommand{\NAOJ}{\affil{National Astronomical Observatory of Japan, 2-21-1 Osawa, Mitaka, Tokyo, 181-8588, Japan}}

\newcommand{\OAN}{\affiliation{Observatorio Astron\'{o}mico Nacional (IGN), C/Alfonso XII, 3, E-28014 Madrid, Spain}}

\newcommand{\Oxford}{\affiliation{Sub-department of Astrophysics, Department of Physics, University of Oxford, Keble Road, Oxford OX1 3RH, UK}}

\newcommand{\Tamkang}{\affiliation{Department of Physics, Tamkang University, No.151, Yingzhuan Rd., Tamsui Dist., New Taipei City 251301, Taiwan}}

\begin{document}

\title{Constraining resolved extragalactic $R_{21}$ variation with well calibrated ALMA observations}

\correspondingauthor{Jakob den Brok}
\author[0000-0002-8760-6157]{Jakob den Brok}
\email{jakob.denbrok@gmail.com}
\CfA

\author[0000-0002-0119-1115]{Elias K. Oakes}
\Connecticut

\author[0000-0002-2545-1700]{Adam~K.~Leroy}
\OSU
\CCAPP

\author[0000-0001-9605-780X]{Eric W. Koch}
\CfA

\author[0000-0003-1242-505X]{Antonio Usero}
\OAN

\author[0000-0002-5204-2259]{Erik~W.~Rosolowsky}
\Alberta

\author[0000-0003-0166-9745]{Frank Bigiel}
\Bonn

\author[0000-0003-0378-4667]{Jiayi~Sun \begin{CJK*}{UTF8}{gbsn}(孙嘉懿)\end{CJK*}}
\altaffiliation{NASA Hubble Fellow}
\Princeton

\author[0000-0001-9020-1858]{Hao He}
\Bonn

\author[0000-0003-0410-4504]{Ashley~T.~Barnes} \ESO

\author[0000-0001-5301-1326]{Yixian Cao}
\MPE

\author[0000-0003-2496-1247]{Fu-Heng Liang}
\ESO
\Oxford

 \author[0000-0002-1370-6964]{Hsi-An Pan}
\Tamkang

\author{Toshiki Saito}
\NAOJ

\author[0000-0002-4781-7291]{Sumit K. Sarbadhicary}
\OSU
\CCAPP

\author[0000-0002-0012-2142]{Thomas G. Williams}
\Oxford


\begin{abstract}

CO(1-0) and CO(2-1) are commonly used as bulk molecular gas tracers. The CO line ratios (especially CO(2-1)/CO(1-0) -- $\rtwo$) vary within and among galaxies, yet previous studies on $\rtwo$ and alike often rely on measurements constructed by combining data from facilities with substantial relative calibration uncertainties that have the same order as physical line ratio variations. Hence robustly determining systematic $\rtwo$ variations is challenging. Here, we compare CO(1-0) and CO(2-1) mapping data from ALMA for 14 nearby galaxies, at a common physical resolution of 1.7\,kpc. Our dataset includes new ALMA (7m+TP) CO(1-0) maps of 12 galaxies. We investigate $\rtwo$ variation to understand its dependence on global galaxy properties, kpc-scale environmental factors, and its correlation with star formation rate (SFR) surface density and metallicity. We find that the galaxy-to-galaxy scatter is 0.05 dex. This is lower than previous studies which reported over 0.1 dex variation, likely reflecting significant flux calibration uncertainties in single-dish surveys. Within individual galaxies, $\rtwo$ has a typical mean value of ${\sim}$0.64 and 0.1 dex variation, with an increase to ${\sim}$0.75 towards galactic centers. We find strong correlations between $\rtwo$ and various galactic parameters, particularly SFR surface density, which shows a power-law slope of 0.10--0.11 depending on the adopted binning/fitting methods. Our findings suggest that, for studies covering main sequence galaxy samples, assuming a fixed $\rtwo$=0.64 does not significantly bias kpc-scale molecular gas mass estimates from CO(2-1). Instead, systematic uncertainties from flux calibration and the CO-to-H$_2$ conversion factor account for more systematic scatter of CO-derived molecular gas properties.

\end{abstract}


%%%%%%%%%%%%%%%%%%%%%%%%%%%%%%%%%%%%%%%%%%%%%%%%%%%%%%%%%%%%%%%%%%%%%
%
%.     Introduction
%
%%%%%%%%%%%%%%%%%%%%%%%%%%%%%%%%%%%%%%%%%%%%%%%%%%%%%%%%%%%%%%%%%%%%%
\section{Introduction} \label{sec:intro}

The lowest rotational-$J$ transitions of carbon monoxide ($^{12}$C$^{16}$O, hereafter CO) constitute an accessible way to trace the bulk distribution and kinematics of the molecular gas across galaxies  \citep{Solomon1987,Young1991, Bolatto2013}. In contrast to the most abundant molecule, H$_2$, the CO molecule has a permanent dipole moment and the ground transition $J=1\rightarrow0$ is excited at ${\sim}5.5\,$K, allowing CO emission to trace gas at the cold temperatures typical of most H$_2$ in galaxies necessary for star formation to occur. %\edit1
{The mass distribution of the molecular gas is then determined using a CO-to-H$_2$ conversion factor, which is expected to be depend on local environment extragalactic environment \citep{Dickman1986, Young1991,Bryant1996, Bolatto2013}.}

Furthermore, the critical density of the low rotational-$J$ CO transitions is relatively modest with %\edit1
{$n_{\rm crit,1-0}{\approx} 570\,\rm cm^{-3}$ and $n_{\rm crit,2-1}{\approx} 4,400\,\rm cm^{-3}$, respectively for the optical thin case at $T_{\rm kin}=20\,K$ \citep[see Table 4 in][]{Schinnerer2024}. For optically thick emission, the critical densities are further reduced by radiative trapping. %, which scale the critical density by the factor $\beta_{ik} = [1-\exp(-\tau)]/\tau$ (where $\tau$ is the optical depth). 
For typical optical depths of $\tau{\sim}5$, the effective critical densities are $n_{\rm crit,1-0} {\approx} 100,\rm cm^{-3}$ and $n_{\rm crit,2-1} {\approx} 500,\rm cm^{-3}$ \citep[consistent with the effective excitation density estimates from][]{Shirley2015}. Most CO emission in galaxies is optically thick, and these effective critical densities align well with the typical mean densities of molecular gas in the ISM \citep{Bolatto2013,Heyer2015}. As a result, both CO(1-0) and (2-1) are effective tracers of the bulk of the molecular gas in galaxies.} Practically, the low rotation-$J$ CO lines are also almost always the lines %\edit1
{that show the highest brightness temperatures in the ground-accessible $\lambda \sim 1{-}3$~mm wavelengths. In contrast, higher-J CO lines (almost always) require longer integration times to be detected and probe a smaller fraction of gas, which is typically warmer and denser than the average, more widely distributed, colder and less dense H$_2$ gas phase}.
%seen from galaxies\footnote{\edit1{We note that this is only true in brightness temperature units.}} 
Given these factors, the lowest rotational-$J$ transitions represent our best tools to observe the molecular gas with ground-based telescopes.

\begin{figure*}
    \centering
    \includegraphics[width=0.95\textwidth]{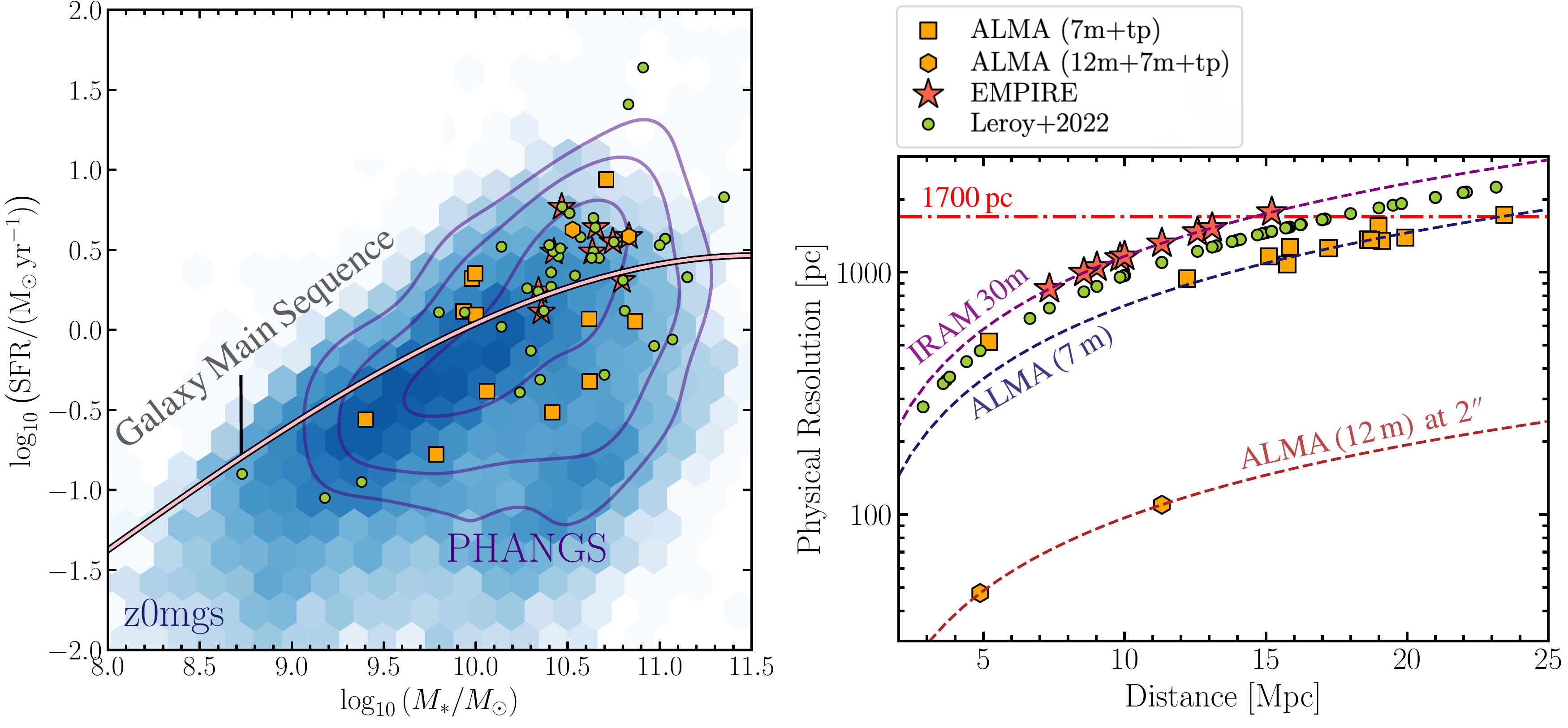}
    \caption{{\bf Sample Overview} ({\textit{Left}}) The panel presents the distribution of studied galaxies in the $\rm SFR{-}M_{*}$ parameter space. The orange squares and hexagons indicate the 14 galaxies tht form our sample. For reference, the red stars present the nine sources analyzed as part of the EMPIRE CO line ratio paper \citep{denBrok2021}. The green circles represent the galaxies for which \citet{Leroy2022} present $R_{21}$ measurements. For context, the contour lines show the 90\%, 75\%, and 33\% inclusion region of PHANGS galaxies in the $\rm SFR{-}M_{*}$ parameter space \citep{Leroy2021_PHANGS}. The blue underlying density distribution illustrates the sample of galaxies within $d{<} 50\rm\,Mpc$ \citep[$N_{\rm gal}{\approx}$11\,000][]{Leroy2019}. The gold and black line illustrates the galaxy main sequence at $z{=}0$ following the parameterization from \citet{Saintonge2017}. ({\textit{Right}}) The panel illustrates the achievable physical resolution based on the observed angular resolution and the galaxy's measured distance. The points follow the description of the left panel. The red, blue, and purple dashed lines indicate the achievable physical resolution as a function of the galaxy's distance for ALMA 12m, ALMA 7m, and IRAM 30m observations. We perform the subsequent analysis at a common physical resolution of 1.7\,kpc, which is illustrated by the red dash-dotted horizontal line. We note that for the ALMA 12m array, multiple configurations are possible that lead to even higher angular (and hence physical) resolutions.}  
    \label{fig:sampledesc}
\end{figure*}

While theoretical work on interpreting and using CO as a H$_2$ tracer often focuses on the CO(1-0) transition \citep[e.g.,][though see \citealp{Gong2020} and \citealp{Penaloza2017}]{Bassini2020,Keating2020,Seifried2020,Hu2023}, observations of galaxies in the local Universe ($z{<}$0.1) have started to target the $J=2\rightarrow1$ transition instead of the $J=1\rightarrow0$ ground transition \citep[e.g.,][]{Leroy2009,Brown2021,Leroy2021_PHANGS,Villanueva2024}. There are several reasons that make \emph{current} observation of CO(2-1) advantageous over CO(1-0): (i) Because CO(2-1) lies at a higher frequency (230.5\,GHz instead of 115.3\,GHz), it can be observed at a higher angular resolution by %\edit1
{a given single-dish telescope or fixed configuration array}; (ii) the CO(1-0) emission line lies 2.7 GHz from the strong O$_2$ telluric absorption feature at 118\,GHz. %\edit1
{Therefore,} under dry atmospheric conditions presents at sites such as the Atacama desert, the CO(2-1) emission is %\edit1
{slightly} less attenuated than the CO(1-0) emission line for targets in the Milky Way or in the nearby Universe (i.e., for small or negligible redshifts of $z{<}0.1$ %\edit1
{as for higher $z$, the line will shift out of the telluric absorption feature}); (iii) when considering the capabilities of the astronomical interferometer ALMA, given (i) and (ii), %\edit1
{for a fixed target angular resolution} the CO(2-1) emission line can generally be mapped 2--4 times faster than CO(1-0) for the same sensitivity and angular resolution\footnote{The reason for this is the high sensitivity of the Band 6 receiver (which covers the CO(2-1) line) and that to reach similar resolution, a configuration with longer baselines needs to be chosen for CO(1-0), leading to a sparser $u-v$ sampling.}. %\edit1
{(iv) CO(2-1) is accessible over a broader redshift range for mm interferometers, enabling studies of molecular gas across a wider range of cosmic time. Given these points, combining high luminosity with the ability to trace the bulk of the H$_2$ mass in GMCs making the CO(2-1) line an important tracer of molecular gas that needs to calibrated for use across diverse environments.}

Given the sensitivity of the CO line intensity to changes in the underlying molecular gas conditions, obtaining accurate measurements of the line ratio $R_{21}$ is crucial for interpreting CO(2-1) observations and translating them into meaningful insights about the molecular gas. %\edit1
{The measured line ratio will depend on the {ISM physical conditions}. In the case of extragalactic observations with large, often $\sim$kpc, beams an individual measurement will combine an ensemble of GMCs with potentially distinct physical {properties (such as size, mass, or average temperature)}.} From theoretical work %\edit1
{on a local, GMC scale level}, we expect the CO emission of different rotational transitions (and hence the resulting CO line ratio $R_{21}${, yet more weakly}) to depend to various degrees on the {GMC-averaged} kinetic gas temperature, $T_{\rm kin}$, the collider density, $n_{\rm H_2}$, the column density of CO, $N_{\rm CO}$, and the CO line velocity dispersion $\Delta v$ \citep{Bolatto2013, Leroy2017dens, Penaloza2017, Gong2020}. As %\edit1
{these local quantities vary as a function of environment} within galaxies, we therefore expect systematic variation of $R_{21}$. 

%\edit1
{$R_{21}$ has been measured and its variations studied in the Milky Way \citep[e.g.,][]{Sakamoto1994, Hasegawa1997, Sawada2001} and nearby galaxies \citep[e.g.,][]{Eckart1990, Braine1992, Braine1993,Papadopoulos1998, Yajima2021, Keenan2024a}. For instance, studying the Orion A and B clouds in the Milky Way, \citet{Sakamoto1994} find and average value of ${\sim}0.66-0.77$, {with values above unity found for the molecular gas near active SF sites, where PDRs and }  H\textsc{ii} {regions are located. Moreover, previous studies have reported}} a correlation of  the ratio $\rtwo$ with the molecular gas density, as $\rtwo$ decreases with lower gas densities from the centers to the edges of Milky Way molecular clouds \citep{Hasegawa1997, Yoda2010}. %\edit1
{In addition, in other galaxies, radial trends have been reported with higher average values toward galactic centers\citep{Braine1992,Leroy2009}.}
Promising tracers of $\rtwo$ variation include the SFR surface density, or the molecular gas depletion time \citep{Narayanan2014,Lamperti2020,denBrok2021, Yajima2021, Leroy2022, Keenan2024a}. However, constraining the variation and obtaining such proxies of the line ratio is not just a technical detail, but a critical step to enable quantitative use of CO(2-1) emission to infer molecular gas masses in nearby galaxies. This, in turn, is key to maximize the scientific return of modern mm-wave observatories.

Extragalactic mapping of CO line ratios have mainly been confined to single- or a few galaxy studies \citep{Crosthwaite2007,Koda2012, Law2018,Koda2020,denBrok2023}. So far, only a small number of studies exist that systematically investigate the CO line ratio across a sample of nearby galaxies \citep{denBrok2021, Yajima2021,Leroy2009,Leroy2013,Leroy2022, Keenan2024a, Keenan2024b}. A major limitation of these studies, however, remains that in the case of single-dish observations, calibration uncertainties can introduce significant scatter in subsequent CO line measurement \citep[see discussion in][]{denBrok2021}. Moreover, if observations from different telescopes are used to compute the CO line ratio, uncertainties in aperture correction constitute an additional point of scatter. Since the dynamical range of $\rtwo$ is small (with values at kpc-scale ranging generally from $0.5-0.9$ in nearby star-forming galaxies), the uncertainties can potentially dominate and hence explain the large scatter reported in the literature. Currently, the interferometer ALMA is the only facility that can provide the the high sensitivity and low calibration uncertainty that are crucial for such a study. The cycle 11 ALMA technical handbook quotes a flux calibration uncertainty of ${\sim}5$\% for band 3 and band 6 observations \citep[see Section 10.2.6 in ][]{Remijan2019}. We discuss the ALMA flux calibration uncertainty in more detail in Section \ref{sec:flux_unc}.  

In this study, we present Cycle 9 ALMA 7m+TP CO(1-0) observation of a set of twelve nearby star-forming galaxies (project code: 2022.1.01479.S; PI: J.~den~Brok). In addition, we include two more archival ALMA CO(1-0) mapping observations: NGC3627 \citep[2015.1.01538.S, PI: R. Paladino; ][]{denBrok2023b} and NGC5236 \citep[2017.1.00079.S, PI: J. Koda; ][]{Koda2023}. In combination with the CO(2-1) ALMA observations as part of the PHANGS survey \citep{Leroy2021_PHANGS}, we have a set of 14 galaxies for which we study the resolved $\rtwo$ at 1.7 kpc physical resolution. The main science questions that we address are:
\begin{enumerate}
     %\item \textit{To what degree is previously observed galaxy-to-galaxy scatter in \rtwo\ driven by \edit2{varying} physical conditions in the gas, and to what extent does this reflect uncertainties relative flux calibrations? }
     \item \textit{To what {degrees do varying physical conditions in the gas, and uncertainties in the relative flux calibration contribute to previously observed galaxy-to-galaxy scatter in \rtwo}?}
     \item \textit{What magnitude of variations does \rtwo\ exhibit at 1.7\,kpc physical resolution contrasting different dynamical environments, including galaxy centers against disks, and spiral arms against interarm regions?}
     \item \textit{How well do local conditions, e.g., the SFR surface density, predict \rtwo\ in nearby galaxies and what are the best current scaling relations to estimate this quantity?}
\end{enumerate}

This paper is structured as follows: In \autoref{sec:obs}, we describe the properties of the sample of fourteen galaxies, the observations and reduction of the new ALMA CO(1-0) data, and the approach used to homogenize the full data set for the science analysis. We describe out methods in \autoref{sec:methods}.  In \autoref{sec:results}, we present the main results, including the overall distribution of $\rtwo$ across all sightlines and the variation of $\rtwo$ within individual galaxies. We discuss the science questions and relate our findings to previous studies in \autoref{sec:disc} and conclude in \autoref{sec:conc}.
%%%%%%%%%%%%%%%%%%%%%%%%%%%%%%%%%%%%%%%%%%%%%%%%%%%%%%%%%%%%%%%%%%%%%
%
%.     Data & Observations
%
%%%%%%%%%%%%%%%%%%%%%%%%%%%%%%%%%%%%%%%%%%%%%%%%%%%%%%%%%%%%%%%%%%%%%

\begin{deluxetable*}{ccccccccccc}

\tablecaption{Sample of Cycle 9 ALMA CO(1-0) observations.\label{tab:obs_summary}}

\tablehead{\colhead{Galaxy} & \colhead{$i$}&\colhead{P.A.}&\colhead{Distance}& \colhead{$\log_{10}\,M_\star$}& \colhead{$\log_{10}\,\rm SFR$}& \colhead{AGN}&\colhead{Beam$^{\rm a}$}&\colhead{$\langle\sigma_{\rm ^{12}CO(1-0)}\rangle^{\rm b}$}&\colhead{$\langle\sigma_{\rm ^{12}CO(2-1)}\rangle^{\rm c}$}&\colhead{Ap. Corr.$^{\rm d}$} \\ 
\colhead{} & \colhead{[deg]}& \colhead{[deg]}& \colhead{[Mpc]} & \colhead{[$M_\odot$]} &\colhead{[$M_\odot\,\rm yr^{-1}$]} &&\colhead{[$''$]}&\colhead{[mK]} &\colhead{[mK]}} 

% All data must appear between the \startdata and \enddata commands
\startdata 
         NGC\,0685& $23.0$ &$100.9$ & 19.94 & 10.06 & -0.38 & \textcolor{red}{\ding{56}}& $17.6$ / $(14.4)$ & $12.9$ / $(17.6)$ & $2.7$ & 1.2\\
         NGC\,1087& $42.9$ & $359.1$ & 15.85 & 9.93  & 0.12&\textcolor{red}{\ding{56}}&$22.1$ / $(16.5)$& $7.4$ / $(11.9)$ & $1.7$ & 1.0\\
         NGC\,1300& $31.8$ &$278.0$ & 18.99 & 10.61	& 0.069&\textcolor{red}{\ding{56}}&$18.5$ / $(16.9)$& $9.7$ / $(12.5)$ & $2.4$& 1.1\\
         NGC\,1317& $23.2$ & $221.5$ & 19.11 & 10.62 & -0.32&\textcolor{red}{\ding{56}}&$18.3$ / $(14.6)$& $11.3$ / $(18.3)$ & $1.7$ & 1.1\\
         NGC\,1385& $44.0$ & $181.3$& 17.22 & 9.98  & 0.32&\textcolor{red}{\ding{56}}&$20.4$ / $(15.1)$& $8.0$ / $(15.1)$ & $1.7$ & 1.0\\
         NGC\,1433& $28.6$ & $199.7$ & 18.63 & 10.87 & 0.055&\textcolor{red}{\ding{56}}&$18.8$ / $(15.1)$& $9.9$ / $(15.5)$ & $3.2$&1.2\\
         NGC\,2566& $48.5$ & $312.0$  & 23.44 & 10.71 & 0.94&\textcolor{red}{\ding{56}}&$15.2$& $16.1$ & $3.9$ & 1.0\\
         NGC\,2835& $41.3$ & $1.0$ & 12.22 & 10.00 &0.095&\textcolor{red}{\ding{56}}&$28.7$ / $(15.9)$& $6.0$ / $(16.2)$ & $2.0$ & 1.4\\
         NGC\,3627$^{*}$& $57.3$ & $173.1$ & 11.32 & 10.83 & 0.58&\textcolor{green}{\ding{52}}&$30.4$ / $(4.5)$&$6.7$ / $(55.7)$ & $1.6$ & 1.3 \\
         NGC\,4457& $17.4$ & $78.7 $ & 15.10 & 10.41 & -0.51	&\textcolor{red}{\ding{56}}&$23.2$ / $(15.9)$& $9.1$ / $(17.7)$ & $2.8$&1.0\\
         NGC\,4540& $28.7$& $12.8 $ &  15.76 & 9.79  & -0.77&\textcolor{red}{\ding{56}}&$22.2$ / $(14.1)$ & $12.1$ / $(27.6)$ & $3.8$ & 1.0\\
         NGC\,5068& $35.7$& $342.4 $ & 5.20  & 9.40	& -0.56	&\textcolor{red}{\ding{56}}&$67.4$ / $(20.5)$& $1.8$ / $(31.4)$ & $1.2$ & 1.3\\
         NGC\,5236$^{*}$& $24.0$ & $225.0 $ & 4.89 & 10.53 & 0.63&\textcolor{red}{\ding{56}}&$71.7$ / $(3.5)$&1.0 / (44.8) & $0.3$ & 1.1\\
         NGC\,7496& $35.9$& $193.7 $ & 18.72 & 10.00 & 0.35	&\textcolor{green}{\ding{52}}&$18.7$ / $(15.0)$& $8.5$ / $(13.0)$ & $2.1$& 1.0\\
\enddata

\tablecomments{(${\rm *}$) Not part of cycle 9 program, but ALMA CO(1-0) and CO(2-1) observations available. $({\rm a})$ This corresponds to the FWHM of the circularized beam for the working resolution (i.e., physical resolution of 1.7\,kpc) and native resolution in parentheses. For NGC\,2566, the working resolution corresponds to the native resolution.  ($\rm b$) The average channel rms at 2.5\,km\,s$^{-1}$ for CO(1-0) at working and native (in parentheses) angular resolution. ($\rm c$) The average rms noise per channel at 2.5\,km\,s$^{-1}$ for PHANGS-ALMA CO(2-1) at working resolution. ($\rm d$) The aperture correction based on the WISE band 3 flux. \\
Sources: Galaxy orientation taken from \citet{Lang2020}; Galaxy distance drawn from \citet{Anand2021}; The galaxy integrated SFR and stellar mass are adopted from \citet{Leroy2022}. }

\end{deluxetable*}

\section{Data and Observations} \label{sec:obs}
\subsection{The sample}
In order to achieve a systematic analysis of the CO line ratio variation across and within a sample of nearby star-forming galaxies, we analyze targets for which ALMA observations exists for both the CO(1-0) and CO(2-1) emission line. A robust benchmark of the CO line ratio $\rtwo$ requires a representative sample of the nearby star-forming galaxy population. This current study is based on a pilot proposal targeting fourteen galaxies drawn from the PHANGS--ALMA survey sample of 90 galaxies for which ALMA 12m CO(2-1) observations exist. The selection for this pilot sample was aimed at ensuring a good dynamic range in galactic SFR and stellar mass. For 12/14 galaxies,  ALMA 7m+TP observations were obtained as part of cycle 9 (project code: 2022.1.01479.S, P.I.: J. den Brok).
In addition, we include two galaxies from the PHANGS sample for which ALMA CO(1-0) and CO(2-1) already exist in the archive. The left panel in \autoref{fig:sampledesc} illustrates the distribution of the selected galaxies in the SFR-$M_{\star}$ parameter space. The figure shows that our targets span more than a factor of $10$ in stellar mass, $M_\star$, and star formation rate, SFR. They  span a smaller but still significant range in specific star formation rate or offset from the main sequence of star forming galaxies. All galaxies in our sample are within a distance of $D{<}25$\,Mpc, resulting in an achievable physical resolution (based on the angular resolution of the ALMA 7m+TP observations) of less than or equal to 1.7\,kpc (see right panel of \autoref{fig:sampledesc}). %The galaxies in our sample are neither edge-on ($i{<}80^\circ$) nor covered by the Milky Way disk ($|b|{>}5^\circ$). 
For reference, Table \ref{tab:obs_summary} lists the galaxy properties of our sample.

\begin{figure*}
    \centering
    \includegraphics[width=0.9\textwidth]{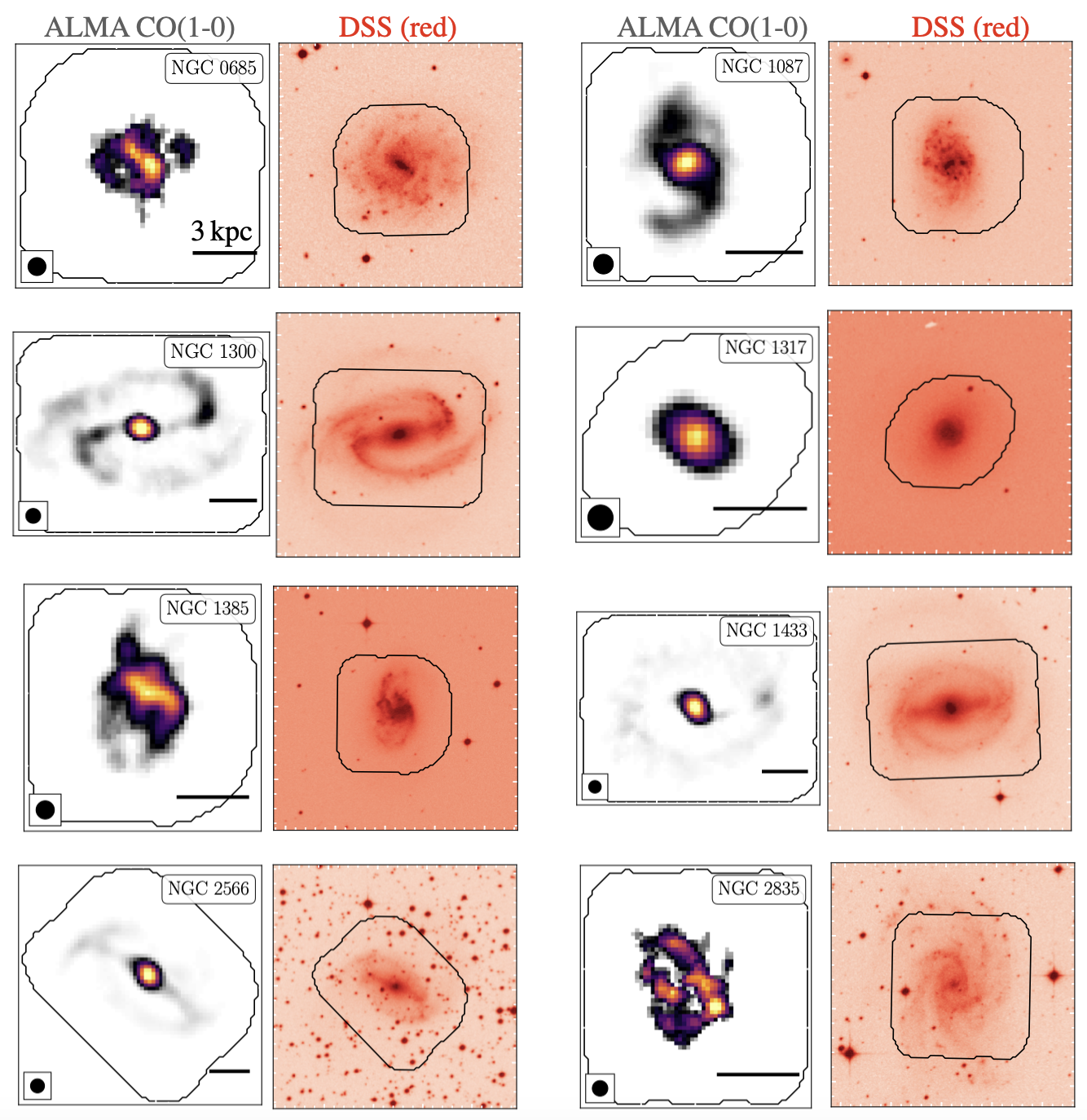}
    \caption{{\bf ALMA CO~(1-0) moment-0 maps} The panel with the galaxy name label in the top right corner present the CO(1-0) observations at the native ALMA 7m resolution (${\sim}15''$). The black circle shows the beam size and the black line indicates a physical scale of 3 kpc, when accounting for the galaxy's distance. For NGC 3627 and NGC 5236, higher angular resolution data is available, but we convolved the maps to 15$''$ for consistency in the presentation. The panels in red to the right illustrate the DSS red maps for reference. The black contour represent the outline of the CO(1-0) ALMA observations. }
    \label{fig:samplemaps}
\end{figure*}

\renewcommand{\thefigure}{\arabic{figure} (Cont.)}
\addtocounter{figure}{-1}

\begin{figure*}
    \centering
    \includegraphics[width=0.9\textwidth]{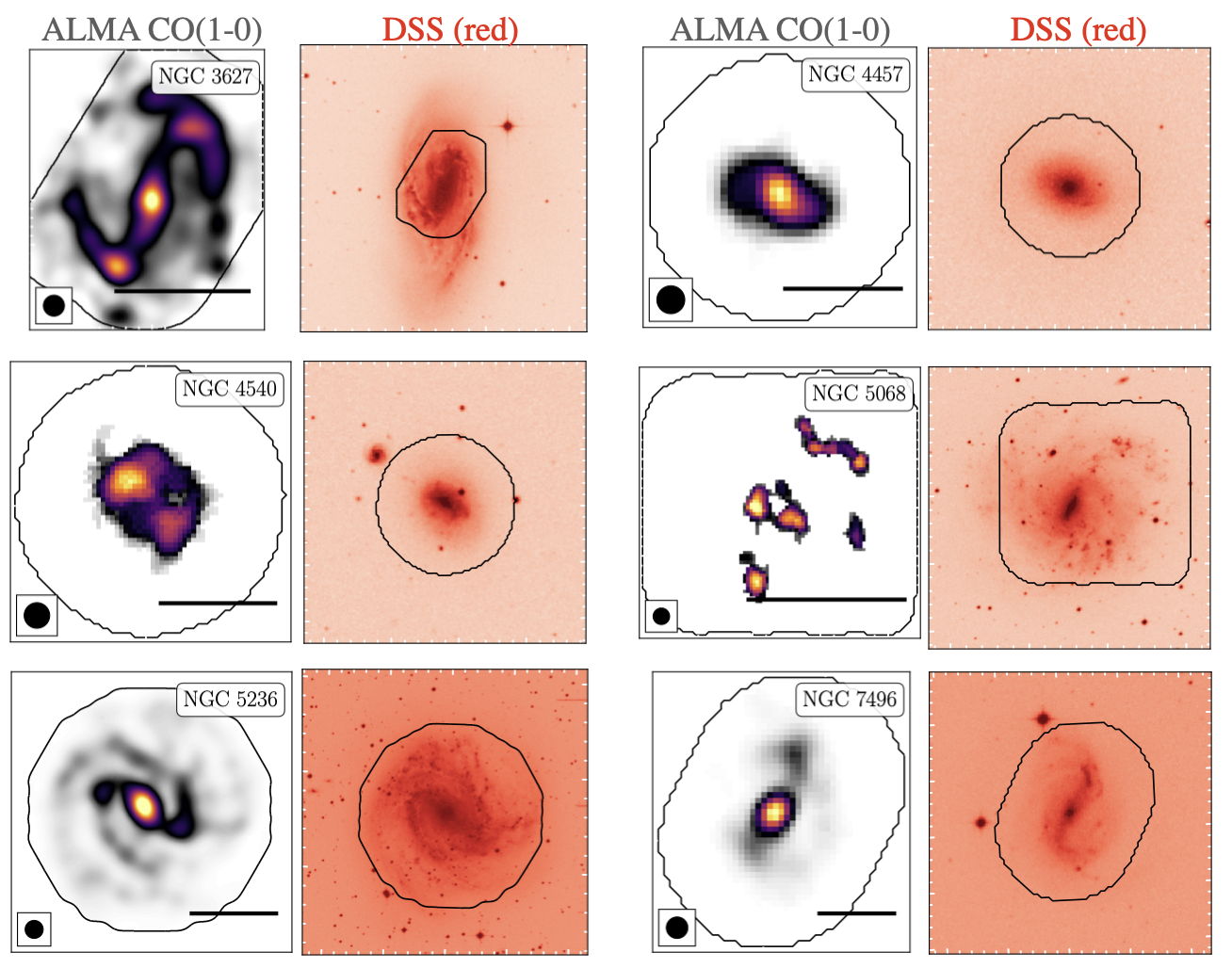}
    \caption{}
    \label{fig:samplemaps_cont}
\end{figure*}

\renewcommand{\thefigure}{\arabic{figure}} 

\subsection{ALMA observations}

\subsubsection{Cycle 9 ALMA 7m and TP observations and data reduction}
This paper presents new Band 3 ALMA CO(1-0) observations of 12 nearby galaxies (Table \ref{tab:obs_summary}). The observations are part of Cycle 9 project 2022.1.01479.S. We observed the CO(1-0) line, rest frequency $115.271$~GHz, using both the total power (TP) and seven meter array (7m) portions of the Morita Atacama Compact Array (ACA). After processing and convolution to a circular synthesized beam, the final observations achieved an angular resolution of approximately 14$''$ to 15$''$. Because we include TP observations, we ensure that no flux is missing. All observations were carried out between September 2022 and November 2023.

We adopted the observatory provided pipeline for the calibration of the CO(1-0) data, which used CASA (v6.4.1) and the ALMA calibration pipeline \citep[][]{Casa2022,Pipeline2023}. Our quality assurance did not reveal any issues with the pipeline, and the data did not require any manual flagging. 

We used the PHANGS-ALMA pipeline\footnote{{v3.1}; \url{https://github.com/akleroy/phangs_imaging_scripts}} to image the calibrated data and produce science-ready data cubes. For a detailed description of the pipeline's functionalities and methods, we refer to the dedicated paper \citet{Leroy2021} or the concise description in Appendix A of \citet{Liu2023}. In short, the imaging pipeline goes through the following steps: (i)  visibility data are converted to a user-defined channel width and frequency range such that the targeted emission line is well sampled; (ii) the continuum is subtracted using a zeroth-order polynomial; (iii) The image is cleaned using repeated the CASA \texttt{tclean} function, emphasizing frequent ``major cycles'' (i.e., frequent re-calculation of the model and residuals in the visibility domain). The routine performs a multi-scale cleaning down to ${\sim}4\sigma$ followed by a single-scale cleaning down to ${\sim}1\sigma$  focused on regions with detected signal. Imaging is considered to converge when successive calls to clean show little influence on the total flux in the model. (iv) Finally, the pipeline performs a series of post-processing steps, which include primary-beam correction, conversion of the cube units to brightness temperature expressed in Kelvin, short-spacing correction using CASA's \texttt{feather} task, and finally convolution of the channels to a common, circularized beam. \autoref{fig:samplemaps} presents an overview of the resulting CO(1-0) moment-0 maps.

In addition to our new observations, our analyses use archival CO(1-0) data sets, Project 2015.1.01538.S (P.I. R. Paladino) targeting NGC~3627 and Project  2017.1.00079.S (P.I J. Koda) targeting NGC~5236 (M83). Both projects include ACA 7m and TP observations, which we process and image in an identical way as the new data. \autoref{fig:samplemaps} also shows these targets convolved to a synthesized beam size of 15$''$ to match our other data. 
%For reference, we show a DSS cutout of the galaxy with the field of view of our observations.

\autoref{tab:obs_summary} lists the achieved CO(1-0) sensitivities at 2.5\,km\,s$^{-1}$ channel width for the working resolution (1.7\,kpc) and at native resolution. At 1.7\,kpc angular resolution, the sensitivity ranges from 1 to 16 mK, with most of the sources (11/14) having a sensitivity of ${<}10$\,mK. Assuming $\alpha_{\rm CO(1-0)}=4.35\,M_\odot\,\rm pc^{-2}/(K\,km\,s^{-1})$, this translates to a $3\sigma$ molecular gas surface density sensitivity of ${\lesssim}{3}\,M_\odot\,{\rm pc^{-2}}$ (assuming a CO(1-0) FWHM of 20\,km\,s$^{-1}$). 
%\textcolor{red}{Comment on the achieved noise here? Ideally also including what surface density you end up approximately sensitive to?}

\subsubsection{PHANGS-ALMA CO(2-1) data}
We combine our CO(1-0) ALMA observations with CO(2-1) data from the PHANGS-ALMA survey \citep{Leroy2021_PHANGS}. The set of CO(2-1) observations include 12m+7m+TP ALMA data, ensuring no missing flux due to lack or undersampling of short baselines. We use first public release (team version 4) of the PHANGS-ALMA data products \citep{Leroy2021}. Though PHANGS-ALMA includes 7m+TP only data products, we use the 12m+7m+TP products and convolve them to our working resolution. We note that the data have been calibrated, processed and imaged using the same approach that we use for the CO(1-0) ALMA data. We expect these to represent the highest quality, most accurate set of CO(2-1) maps provided by the survey. 
%Note, though that these CO(2-1) data products have median synthesized beam of $1''.3$, much smaller synthesized beam than our working resolution of ${\sim}15''$. 
After convolution to matched resolution, the PHANGS-ALMA CO(2-1) data have typical noises of ${<}4\,$mK at 2.5\,km\,s$^{-1}$ channel width. Therefore, the CO(2-1) data surpasses the CO(1-0) data in terms of sensitivity by a factor 2--4 for each galaxy (for a comparison we refer to \autoref{tab:obs_summary}). 
%\textcolor{red}{add!. Then comment on contrast with the CO(1-0).}

\subsubsection{ALMA calibration uncertainty and total flux recovery}
\label{sec:flux_unc}
A key requirement for successfully benchmarking the $R^{12}_{21}$ variation is a %\edit1
{robust flux calibration with low uncertainty}. ALMA regularly monitors quasars and hence can achieve a total power calibration with gain uncertainties at Band 6 frequencies around 5\%–10\% and Band 3 frequencies around 5\% as stated in the ALMA technical notes \citep[see Section 10.2.6 in ][]{Remijan2019}. 
%These uncertainties also agree with the analysis in \citep{Leroy2021}, who report a ${\sim}3\%$ for the total power variation between different execution blocks (i.e. observations).
%\edit1
{Analyzing the TP observations for PHANGS--ALMA, including the galaxies that we study here, \citet{Leroy2021} report a low rms of $2-4\%$ rms variation of the intensity among repeat observations of a given high S/N source across different dates. They also showed good overall agreement between the PHANGS--ALMA CO~(2-1) maps and previous IRAM 30-m mapping, with the cases of poor agreement attributable to known issues with the IRAM 30-m data \citep[see also][]{denBrok2021}.}
\citet{Francis2020} report even lower flux calibration uncertainties, with only ${\sim}1\%$ for Band 3 observations for a carefully selected spectral setup. Therefore, we infer a conservative estimate of a ${\sim}5\%$ flux calibration uncertainty for the ALMA data. %\edit1
{We note that this 5\% uncertainty is significantly lower than the calibration uncertainty reported for IRAM 30m CO mapping observations using the EMIR (10\%-15\%; \citealt{denBrok2021}) or HERA (${\lesssim}20\%$ based on a detailed analysis in \citealt{Leroy2009,denBrok2021}) receivers. This is also much less than the uncertainty associated with the Nobeyama 45m COMING survey, for which \citet{Yajima2021} cite a combined calibration and pointing uncertainty of 25\%.}

%\edit1
{ALMA is constructed with the goal of recovering the true sky intensity via the inclusion of 7-m and TP observations. However, all astrophysical interferometers have some incomplete $u-v$ coverage, and this will have an effect on the final flux recovery. For PHANGS--ALMA, \citet{Leroy2021} and \citet{Neumann2023_stacking} gauged the uncertainty associated of $u-v$ sampling and image reconstruction on the output image by running simulated galaxy images through the PHANGS---ALMA data processing pipeline \citep{Leroy2021} and then applying spectral stacking techniques \cite{Neumann2023_stacking}. They found that when the TP data are included, the majority of the signal of mm line emission across galaxies is recovered by the 7m+TP observations within 10\% and often much better. Our conclusions still hold even when assuming a higher uncertainty of ${\sim}10\%$. } {For reference only, given the detailed investigations by \citet{Leroy2021} and \citet{Neumann2023_stacking}, we show the overlap in $u-v$ space for CO(1-0) and CO(2-1) for one galaxy in our sample in Appendix \ref{app:flux_recov} to demonstrate that we have sufficient overlap to recover the CO fluxes.}

\subsection{Ancillary data sets}
In this study, we require additional multi-wavelength observations to correlate the CO line ratio with various galactic properties. Here we briefly describe the origin and products of the key supplementary data sets. 

\subsubsection{SFR and stellar mass surface density}
In order to obtain matched-resolution star-formation and stellar mass surface density estimates, we use a combination of narrowband H$\alpha$ and Wide-field Infrared Survey Explorer (WISE) 22 $\mu$m data. The narrowband H$\alpha$ have been acquired with the 2.5 m du Pont
telescope and the ESO/MPG 2.2 m telescope (see A. Razza
et al. 2024, in preparation). And the WISE data are part of the z0mgs survey products \citep{Leroy2019}. The combination of the optical and infrared flux and subsequent conversion to a star formation ratio follows the new calibration presented in \citet{Belfiore2023}. This prescription is better adapted at minimizing the contamination from the infrared cirrus in the 22 $\mu$m band than the commonly used \citet{Calzetti2007} prescription.

The stellar mass surface density, $\Sigma_{*}$, is estimated using Spitzer IRAC 3.6$\mu$m near-infrared maps \citep[][most from S4G \citealt{Sheth2010}]{Fazio2004}. We refer to \citet{Leroy2021_PHANGS} for a detailed description of the approach used to estimate the stellar mass surface density. In short, the NIR image is  converted to a stellar mass surface density using a mass-to-light ratio that depends on an observed color related to the local specific star formation rate. In practice, the mass-to-light ratio is calculated based on the ratio of the SFR and the WISE1 (3.4$\mu$m) band using a scaling relationship calculated by \citet{Leroy2019} to match the results of population synthesis modeling of the SDSS main galaxy sample by \citet{Salim2016,Salim2018}. That modeling adopted the \citet{Bruzual2003} stellar population models and assumes a \citet{Chabrier2003} IMF. The resulting stellar mass estimates are overall consistent with population synthesis model of the PHANGS-MUSE data and are consistent with previous studies, such as \citet{Leroy2022}.

\subsubsection{Metallicity}
For 8 of our 14 targets, we use metallicity measurements obtained from PHANGS-MUSE observations \citep{Emsellem2022}
%For 8/14 galaxies,
\footnote{The sample of galaxies for which we have MUSE-IFU observations are: NGC1087, NGC1300, NGC1385, NGC1433, NGC2835, NGC3627, NGC5068, and NGC7496.}. For the analysis related to the metallicity, we therefore drop the remaining six galaxies and only focus on the sample of 8 for which measurements are available.
%we can employ azimuthally varying metallicity measurements. The 2D distribution of metals have been calculated using a Gaussian Process Regression. 
We specifically use the metallicity maps from  \citet{Williams2022}. The 2D metallicity distribution are determined using a Gaussian process regression by extrapolating measurements of individual pointings within the galaxies and the radial trends. The metallicities for individual pointings are estimated using the S-calibration system of \citet{Pilyugin2016}. See \citet{Groves2023} for more discussion of the metallicity estimates for PHANGS-MUSE.
%In the analysis 
%for details on the method. In short, using the MUSE-IFU observations, the metallicity  is estimated with the S-calibration \citep{Pilyugin2016} for H\textsc{ii} regions throughout the galaxies. 
%From this discrete sampling of sightlines across the galaxies, a 2D map is then computed using a Gaussian process regression.

\subsection{Homogenizing and processing the data} \label{sec:pystruct}
To ensure a systematic analysis, we convolve the cubes to a common physical resolution of 1.7\,kpc using a Gaussian beamsize and regrid to a hexagonal grid. We use the publicly available \texttt{PyStructure} pipeline\footnote{\url{https://github.com/jdenbrok/PyStructure}} \citep{PyStructure_v3}, which has been employed in previous studies \citep[e.g.,][]{denBrok2021,Eibensteiner2022,Neumann2023}.
This versatile tool allows the combination of a set of different 3D data cubes and 2D maps from different telescopes and at different frequencies into a common data structure. All data are then convolved to a user-defined angular (or physical) resolution and regridded to the same grid using a user-defined prior cube. The points of the resulting grid are half-beam spaced in a hexagonal pattern. This pattern optimizes the oversampling with respect to the commonly used Cartesian grid. The pipeline also deprojects the coordinates using the position angle, inclination, and distance of the individual galaxies. This way, we can compute the galactocentric radius of each sightline as well as the Vaucouleurs radius ($r_{25}$). %For the sources in our study, we use the orientation and distances taken from \citet{Lang2020} and \citet{Anand2021} respectively.

The pipeline then performs a postprocessing of the 3D data cubes and computes the moment maps. To integrate, the tool first determines a velocity-range based on a user-defined prior line. In our case, we select the CO(2-1) emission line due to its higher S/N than the CO(1-0).  The signal mask over the entire cube is then determined using a method adapted from \citet{Leroy2021}. For the creation of the signal mask, first, the noise is determined along both the spatial and spectral dimensions over the signal-free part of the cube using the \texttt{mad\_std} function from \texttt{astropy}. This function computes the median absolute deviation. A factor of 1.4826 is then applied to convert it into a standard deviation assuming a Gaussian distribution of the noise.  Then the pipeline first selects all voxels that satisfies the high ($4\sigma$) S/N  threshold. In a next step, the resulting signal mask is expanded to include all adjacent voxels that are contained within the lower ($2\sigma$) S/N mask. Finally, the signal mask is extended along the velocity axis by another 10 km\,s$^{-1}$.

The resulting signal mask is then applied to compute the peak intensity, the integrated intensity (moment-0), the velocity weighted mean (moment-1), and the velocity dispersion (in terms of FWHM). We note that the moment-1 and moment-2 calculations are sensitive to noise effects. Therefore, an additional signal mask is computed for each line separately for the purpose of the moment-1 and moment-2 per line only. This mask is hence not based on the brighter or higher-S/N prior line. The pipeline also computes the uncertainty of the moment maps by propagating the rms it determines from the signal free part of the spectrum. The precise calculation follows the prescriptions from \citet{Leroy2021}.

We also provide aperture correction values for the global (galaxy-wide) measurements as our field of view of the CO(1-0) and CO(2-1) observations generally do not cover the entire disk of the galaxy, but are limited to the region active with star formation. We expect to cover most of the CO flux within our observed field of view, but note that fainter CO emission might extent into the outer disks of galaxies \citep[see for example ][]{Young1995, Braine2007, Schruba2011}.  To quantify the corresponding \emph{aperture} correction we follow the approach outlined in \citet{Leroy2021}. The method consists taking the ratio of the  WISE band 3 flux summed over the entire galaxy with repsect to the flux within the field-of-view of our ALMA observations.
The motivation for taking the band WISE 3 data is that a close correlation exists with the CO integrated intensity \citep[see for more details ][]{Leroy2021}.

\section{Line Ratio Measurements}
\label{sec:methods}

\subsection{Resolved CO line intensity ratio}
We measure the ratio for each 1.7 kpc physical scale resolved line of sight. In particular, we define the integrated line intensity ratio as
\begin{equation}
    \rtwo\equiv \rtwo^{\rm 1.7\,kpc}=\frac{W_{\rm CO}^{2\rightarrow1}}{W_{\rm CO}^{1\rightarrow0}}
\end{equation}
where $W_{\rm CO}^{2\rightarrow1}$ and $W_{\rm CO}^{1\rightarrow0}$ represent the moment-0 value as extracted from the \texttt{PyStructure} (see \autoref{sec:pystruct}) in units of $\rm K\,km\,s^{-1}$ at 1.7\,kpc physical resolution. For reference, we note that in our line ratio framework the line ratio for an optically-thick and thermalized line will tend toward approximately unity\footnote{As described in \citet{Leroy2022}, cosmic microwave background effects and the Raleigh-Jeans approximation will result in slight deviations such that the thermalized line ratio will not exactly tend toward unity.} \citep{Solomon2005}. 
%In the following study, 
When presenting results for individual sightlines and histograms or maps maps of the ratio, we include those sightlines for which both the CO(1-0) and CO(2-1) integrated intensities exceed a S/N${>}$5. We note that for all galaxies in our sample, the CO(2-1) is observed at higher sensitivity, so generally, the CO(1-0) is the limiting factor for whether a sightline is classified as significant (i.e., S/N${>}$5 for both lines) or not (see \autoref{tab:obs_summary}). %Therefore, the limiting factor for a significant $\rtwo$ measurement is our ability to detect CO(1-0). 
Our signal clipped mask at S/N${>}$5 still contains the majority of the CO emission over the full galaxy. We compare the flux within the S/N${>}$5 mask to the flux across the entire map. We find a median completeness of the ALMA data across the sample of 98\% for CO(1-0) and 96\% for CO(2-1) (see \autoref{table:results_ratio} for completeness values per galaxy).
%For reference, we provide the completeness fraction per galaxy and number of individually detected sightlines in \autoref{table:results_ratio}.
%\textcolor{red}{Probably need to present some quantitative statement on which data set limits this and how low you go in intensity. I would consider strongly to explicitly quote ``completeness'' i.e., fraction of flux included in S/N>5 detections (perhaps in Table 1) to make clear how much goes undetected.} 
We use the Gaussian error propagation to derive an estimate of the CO ratio uncertainty from the uncertainty of the individual integrated intensities, which incorporates the channel rms and a flux calibration uncertainty of 5\% (see Section \ref{sec:flux_unc}).

\subsection{Calculating an average CO line ratio}\label{sec:mean}

A key result of this study are measurements of the average $\rtwo$ line ratio across or within certain parts of the galaxies and quantification of the corresponding scatter. We report the average value calculated through three different methods: 

\paragraph{$\langle \rtwo\rangle$ -- The resolved line ratio average} This measurement corresponds to the area weighted median, which means that we weight each sightline per bin equally. We describe the scatter taking the 16$^{\rm th}$-to-50$^{\rm th}$ percentile as a $1\sigma$ interval below the median, corresponding to the quantiles of Gaussian distribution.  Similarly, the 
50$^{\rm th}$-to-84$^{\rm th}$ percentile range is a $1\sigma$ interval above the median. We note that for this measurement, we only average over significantly detected $\rtwo$ values. 

\paragraph{$\langle\rtwo\rangle_{\Sigma}$ -- The ratio-of-sums} which we obtain by summing over pixels of CO(1-0) and CO(2-1) {integrated} intensities and then taking the ratio of the two (following definition in \citealt{Leroy2022}). In the case where the area that we sum over covers the whole galaxy, this value would correspond to an unresolved galaxy measurement. In practice, we sum over all pixels where there is measured CO emission for both lines (see description of how we construct such a signal mask in Section \ref{sec:pystruct}). 
For this approach we do not provide a percentile range but propagate the uncertainties of the individual intensities we sum over.

\paragraph{$\langle\rtwo\rangle_{\rm stacked}$ -- Spectrally stacked ratio}
We also employ a spectral stacking method to improve the S/N and detect potential trends of $\rtwo$ with various physical quantities.
The spectral stacking method resembles the one detailed in \citet{Schruba2011} and its application and robustness for interferometric data that include short spacing data has been documented in \citet{Neumann2023_stacking}. We use the tool \texttt{PyStacker}\footnote{\url{https://github.com/PhangsTeam/PyStacker}} to perform the spectral stacking.
In short, the technique consists of binning sighlines by a parameter of interest. Each spectrum for each sightline is regridded along the velocity axis such that the local mean velocity corresponds to 0 km\,s$^{-1}$. This allows us to coherently add spectra from different locations in the galaxy or even different galaxies within a bin. We average together all spectra in each bin along each channel to obtain a stacked spectrum for both CO(1-0) and CO(2-1). %Finally, the code computes a signal mask using a user-defined prior line (the line with the highest S/N). The mask is computed by determining all channels above 4$\sigma$ and then expanding the channels to the side down to 2$\sigma$.
Both lines are integrated over the same (moment-0) mask and then their ratio forms the stacked $\rtwo$ measurement. 
In contrast to the binning of only significant detections described above, we apply this stacking to all sightlines. The difference between the two sets of results indicate how much faint undetected emission contributes to the overall results. 
%\edit1
{Since a new mask over which to integrate is computed for each stacked spectral bin, the resulting ratio might vary slightly with resect to the \emph{ratio-of-sums}, even in the case where all lines of sight within the bins have significant detected CO emission.}

\subsection{Binning CO line ratios}

To highlight trends for the resolved sightlines across or within our sample of galaxies we employ a binning analysis. For each parameter of interest, we define a set of bins (e.g., of $\Sigma_{\rm SFR}$, $r_{\rm gal}$, etc.) and group together all sightlines according to those bins. Then, within each bin we perform the calculations of the average line ratios described in Section \ref{sec:mean}. We note that for the average ratio ($\langle R_{21}\rangle$), we only bin significant measurements, where both lines are detected at 5$\sigma$. In contrast, for the ratio of sums and stacking, we also included non-detections (i.e., ${<}5\sigma$) when binning.

%\paragraph{Quantities used for binning} 
We bin by four quantities: (i) galactocentric radius, $r_{\rm gal}$.
% in units of kpc
%and in terms of the Vaucouleur radius ($r_{25}$). 
%In kpc units, the bins have a width of 1kpc. %, and in $r_{25}$ we use bins of 0.1 width, in both cases starting from 0. 
(ii) The SFR surface density, $\Sigma_{\rm SFR}$. We  bin by the value in units of $M_\odot\,\rm yr^{-1}$~kpc$^{-2}$ directly. %, or we normalize the SFR surface density per galaxy by the galaxy-wide average, $\langle\Sigma_{\rm SFR}\rangle_{\rm gal}$.
(iii) The specific SFR, which is given by the ratio $\Sigma_{\rm SFR}/\Sigma_{\rm *}$. %Here again we also compute bins and trends for the value directly and for the normalized value by the galaxy-wide averages.
(iv) The metallicity $Z$ in terms of $12+\log\left(\rm O/H\right)$.  We note that we choose to bin by quantities that do not involve CO-derived values to avoid correlated axes.

%%%%%%%%%%%%%%%%%%%%%%%%%%%%%%%%%%%%%%%%%%%%%%%%%%%%%%%%%%%%%%%%%%%%%
%
%.     Results 
%
%%%%%%%%%%%%%%%%%%%%%%%%%%%%%%%%%%%%%%%%%%%%%%%%%%%%%%%%%%%%%%%%%%%%%

\begin{figure*}
    \centering
    \includegraphics[width=0.95\textwidth]{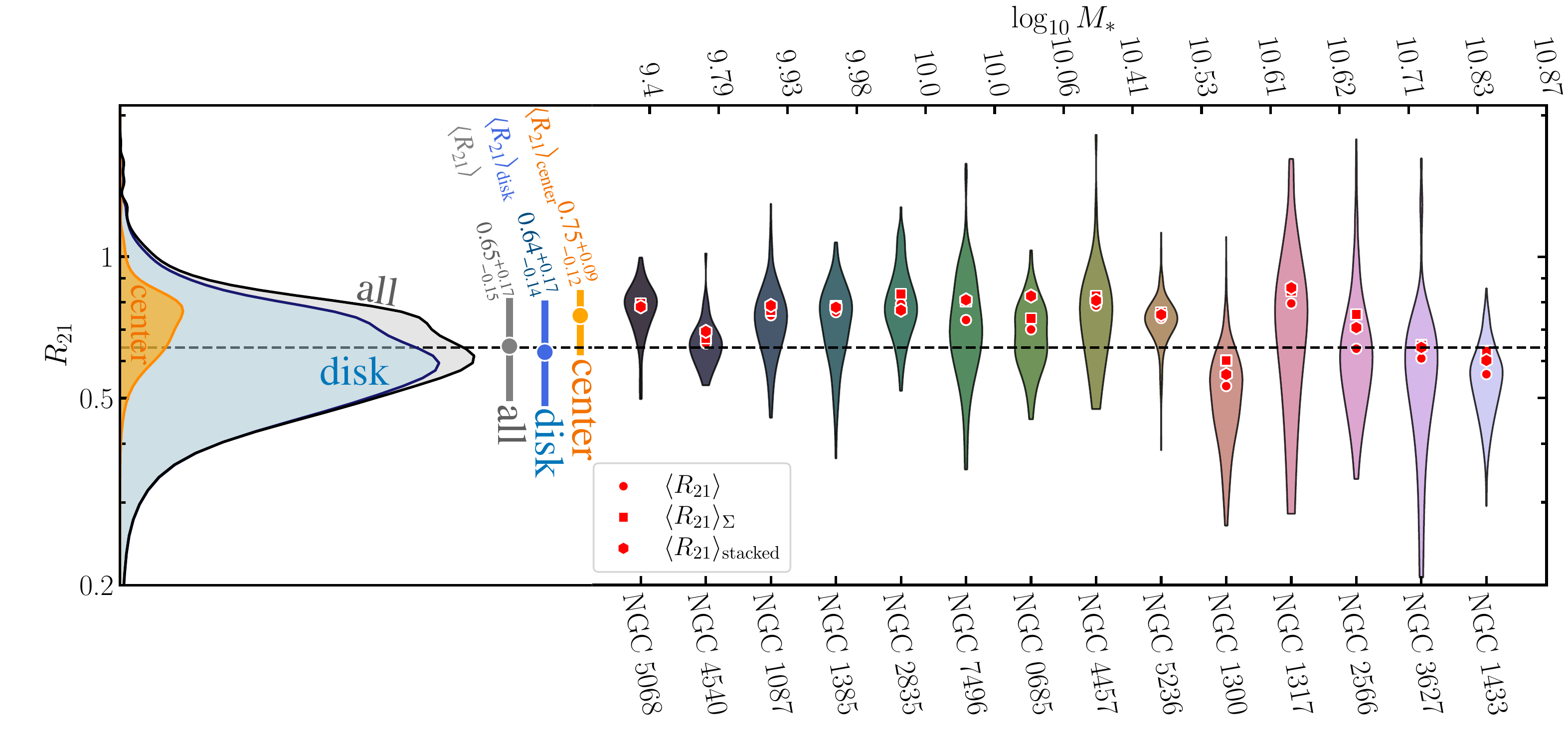}
    \caption{{\bf Sample-wide \rtwo\ distribution}   ({\textit{Left}}) The kernel density estimation (KDE) of all significantly detected sightlines across the full galaxy sample. Here we also separate by sightlines associated with center (orange shaded region); disk (cyan shaded region). For reference, we indicate the equally weighted (i.e., area weighted) mean and 16$^{\rm th}$-to-84$^{\rm th}$ percentile range of the three different KDE's in the top. (\textit{Right}) These violin plots show the distribution of \rtwo\ for sightlines with significant CO(2-1) and CO(1-0) detections for each galaxy. We illustrate the area-weighted median with a circle, the ratio-of-sum average with a square, and the stacked average with a hexagon marker. The dashed line indicates the sample-wide average when weighting all sightlines with significant line ratio detections equally.}
    \label{fig:r21_distr_overall}
\end{figure*}

\begin{figure*}
    \centering
    \includegraphics[width=0.95\textwidth]{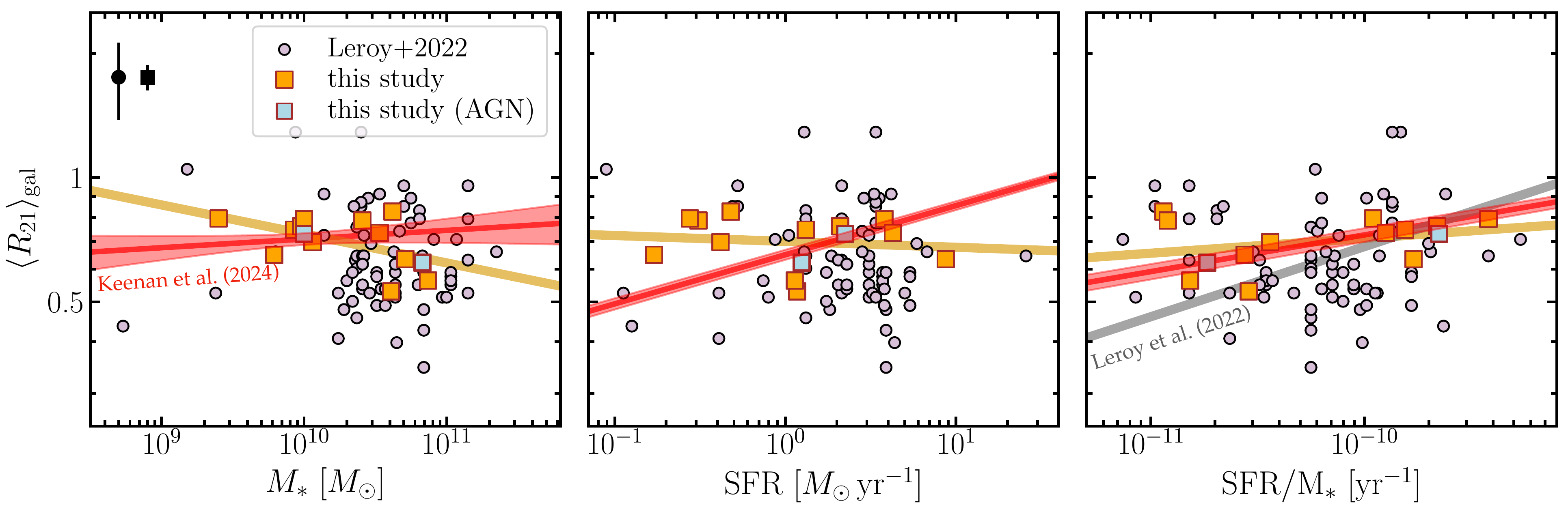}
    \caption{{\bf Trends of galaxy-wide $R_{21}$ and galaxy-integrated properties} These panels illustrate the equally-weighted median $\rtwo$ per galaxy with respect to the galaxy-wide stellar mass, SFR and specific SFR. For reference, we have included the sample of galaxies studied by \citet{Leroy2022}. We perform a power-law regression to our sample of fourteen galaxies, which is indicated with the orange line. Two of the sources in our sample host an AGN (NGC3627 and NGC7496; highlighted in blue). The typical error (dominated by calibration uncertainties) is illustrated in the top left corner of the left panel. %\edit1
    {For reference, we also show the scaling relation fit reported by \citet[][shown in red]{Keenan2024b} and \citet[][shown in grey, and is available only for the $\rm SFR/M_\star$]{Leroy2022}.} }
    \label{fig:r21_galToGal_trends}
\end{figure*}

\begin{figure}
    \centering
    \includegraphics[width=0.9\columnwidth]{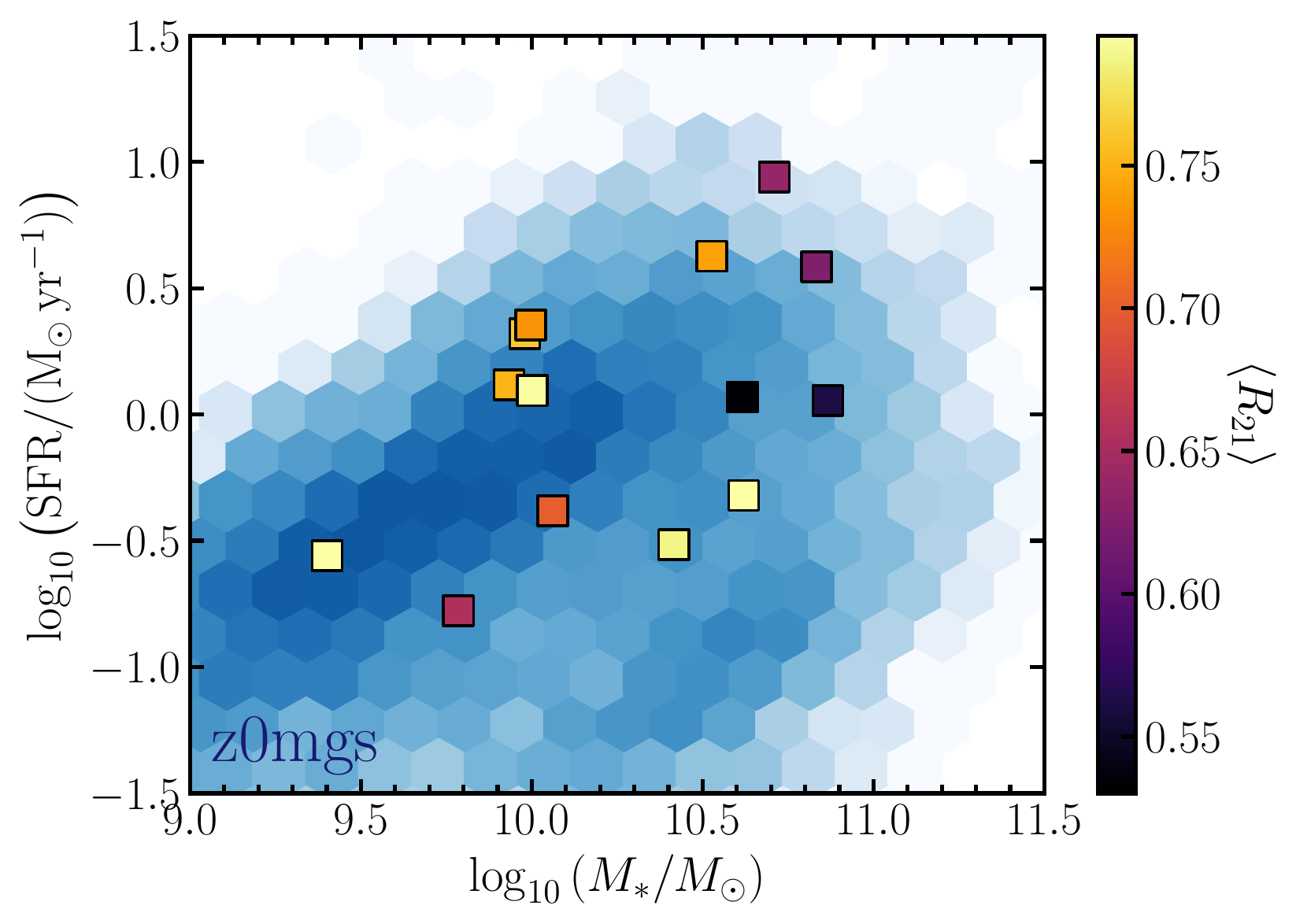}
    \caption{{\bf Distribution of galaxy-integrated $\langle \rtwo\rangle$ in the SFR-M$_*$ parameter space} The 14 galaxies in our sample are colored based on the galaxy-wide average $\rtwo$ (see \autoref{table:results_ratio} for values). As reference, we illustrate also the z0mgs-sample distribution (see \autoref{fig:sampledesc}).}
    \label{fig:r21_SFRMstar}
\end{figure}

\begin{deluxetable*}{l c c c c c c | c c}
\tablewidth{0pt}
\tablecaption{ Summary of line ratio results (highlighted in bold is the recommended value to use for a sample of galaxies). \label{table:results_ratio}}
\tablehead{
&\multicolumn{6}{c}{Resolved}&\multicolumn{2}{c}{Unresolved} \\
 & & & & \multicolumn{3}{c}{$\langle \rtwo\rangle$} &    \\ \colhead{Galaxy} & $n_{\rm pixel}$ & $f^{\rm{\ge}5S/N}_{\rm CO(1-0)}$ & $f^{\rm{\ge}5S/N}_{\rm CO(2-1)}$  & \colhead{entire map}  & \colhead{center} & \colhead{disk} &\colhead{$\langle \rtwo\rangle_\Sigma$} &\colhead{$\langle \rtwo\rangle_{\rm stacked}$}
}
\startdata
NGC 0685 & $71$ &  80\% & 76 \% & $0.70^{+0.10}_{-0.10}$& $0.67^{+0.10}_{-0.13}$& $0.70^{+0.11}_{-0.10}$& $0.74^{+0.05}_{-0.05}$& $0.82^{+0.06}_{-0.06}$\\ [5pt]
NGC 1087 & $116$ &  98\% & 98 \% & $0.75^{+0.10}_{-0.10}$& $0.78^{+0.04}_{-0.05}$& $0.73^{+0.13}_{-0.10}$& $0.76^{+0.05}_{-0.05}$& $0.78^{+0.06}_{-0.06}$\\ [5pt]
NGC 1300 & $343$ &  98\% & 97 \% & $0.53^{+0.09}_{-0.11}$& $0.63^{+0.13}_{-0.06}$& $0.52^{+0.09}_{-0.10}$& $0.56^{+0.04}_{-0.04}$& $0.55^{+0.04}_{-0.04}$\\ [5pt]
NGC 1317 & $39$ &  99\% & 98 \% & $0.83^{+0.29}_{-0.24}$& $0.83^{+0.22}_{-0.13}$& $0.69^{+0.46}_{-0.22}$& $0.83^{+0.06}_{-0.06}$& $0.85^{+0.06}_{-0.06}$\\ [5pt]
NGC 1385 & $128$ &  98\% & 98 \% & $0.76^{+0.11}_{-0.13}$& $0.77^{+0.03}_{-0.04}$& $0.75^{+0.13}_{-0.13}$& $0.78^{+0.06}_{-0.06}$& $0.78^{+0.06}_{-0.06}$\\ [5pt]
NGC 1433 & $309$ &  97\% & 96 \% & $0.56^{+0.09}_{-0.09}$& $0.65^{+0.10}_{-0.09}$& $0.56^{+0.09}_{-0.09}$& $0.61^{+0.04}_{-0.04}$& $0.60^{+0.04}_{-0.04}$\\ [5pt]
NGC 2566 & $277$ &  99\% & 98 \% & $0.63^{+0.22}_{-0.13}$& $0.85^{+0.46}_{-0.21}$& $0.63^{+0.19}_{-0.13}$& $0.73^{+0.05}_{-0.05}$& $0.71^{+0.05}_{-0.05}$\\ [5pt]
NGC 2835 & $84$ &  97\% & 95 \% & $0.79^{+0.17}_{-0.09}$& $0.77^{+0.07}_{-0.10}$& $0.81^{+0.19}_{-0.09}$& $0.81^{+0.06}_{-0.06}$& $0.78^{+0.06}_{-0.06}$\\ [5pt]
NGC 3627 & $97$ &  99\% & 98 \% & $0.62^{+0.11}_{-0.13}$& $0.71^{+0.04}_{-0.05}$& $0.60^{+0.12}_{-0.12}$& $0.64^{+0.05}_{-0.05}$& $0.63^{+0.04}_{-0.04}$\\ [5pt]
NGC 4457 & $37$ &  99\% & 99 \% & $0.79^{+0.13}_{-0.16}$& $0.81^{+0.13}_{-0.10}$& $0.72^{+0.13}_{-0.14}$& $0.82^{+0.06}_{-0.06}$& $0.80^{+0.06}_{-0.06}$\\ [5pt]
NGC 4540 & $36$ &  100\% & 100 \% & $0.65^{+0.03}_{-0.06}$& $0.65^{+0.03}_{-0.04}$& $0.65^{+0.04}_{-0.07}$& $0.65^{+0.05}_{-0.05}$& $0.67^{+0.05}_{-0.05}$\\ [5pt]
NGC 5068 & $25$ &  100\% & 100 \% & $0.80^{+0.04}_{-0.07}$& $0.80^{+0.03}_{-0.05}$& $0.80^{+0.04}_{-0.07}$& $0.79^{+0.06}_{-0.06}$& $0.78^{+0.06}_{-0.06}$\\ [5pt]
NGC 5236 & $132$ &  100\% & 100 \% & $0.73^{+0.04}_{-0.05}$& $0.79^{+0.02}_{-0.04}$& $0.73^{+0.04}_{-0.05}$& $0.76^{+0.05}_{-0.05}$& $0.76^{+0.05}_{-0.05}$\\ [5pt]
NGC 7496 & $118$ &  98\% & 98 \% & $0.73^{+0.20}_{-0.13}$& $0.68^{+0.22}_{-0.03}$& $0.73^{+0.20}_{-0.14}$& $0.79^{+0.06}_{-0.06}$& $0.80^{+0.06}_{-0.06}$\\\hline
sample-wide & $1812$ &  99\% & 98 \% & $\boldsymbol{0.64^{+0.16}_{-0.14}}$& $0.75^{+0.09}_{-0.12}$& $0.63^{+0.17}_{-0.14}$& $0.70^{+0.05}_{-0.05}$& $0.71^{+0.05}_{-0.05}$\\
\enddata
\tablecomments{For the resolved ratios, we list the median and 16$^{\rm th}$ to 84$^{\rm th}$ percentile range. }
\end{deluxetable*}

\section{Results} \label{sec:results}
\subsection{The sample-wide \texorpdfstring{$R_{21}$}{Lg} distribution}

In total, our post-processed data set consists of 1812 half-beam sampled sightlines\footnote{Given an oversampling factor of ${\sim}4$, this corresponds to approximately $\sim$400 independent data points} across the sample of 14 galaxies for which a significant $\rtwo$ exists (i.e. $\rm S/N{>}5$). %\textcolor{red}{See comment, suggest to give details on flux including and what limits detections here.} 

\subsubsection{Galaxy-scale averages}
The left panel in \autoref{fig:r21_distr_overall} illustrates the overall distribution of significantly measured line ratios and we report the corresponding statistics in \autoref{table:results_ratio}. 
We find an area weighted mean and 16$^{\rm th}$-to-84$^{\rm th}$ percentile range across all sightlines of $\langle \rtwo\rangle=0.64^{+0.16}_{-0.14}$. When we take the ratio of the sums of intensities across the galaxy and the stacked ratio, we get a higher average of $\langle \rtwo\rangle_{\Sigma}=0.70^{+0.05}_{-0.05}$ and $\langle \rtwo\rangle_{\rm stacked}=0.71^{+0.05}_{-0.05}$, respectively. The higher value of the average $\langle\rtwo\rangle$ likely reflects the fact that these averages constitute a intensity weighting and sightlines toward galaxy centers, which are intrinsically bright but represent few total sightlines, tend to have higher \rtwo . This is also highlighted in the right panel of \autoref{fig:r21_distr_overall} where the distribution of all sightlines is separated into disk and center. Here the center is defined as sightlines with $r_{\rm gal}{\le}2\,\rm kpc$. Indeed, if only selecting the center sightlines, the average ratio (weighting all central sightlines equally) is even higher with $\langle \rtwo\rangle_{\rm center}=0.75^{+0.09}_{-0.12}$. 

\begin{figure*}
    \centering
    \includegraphics[width=0.85\textwidth]{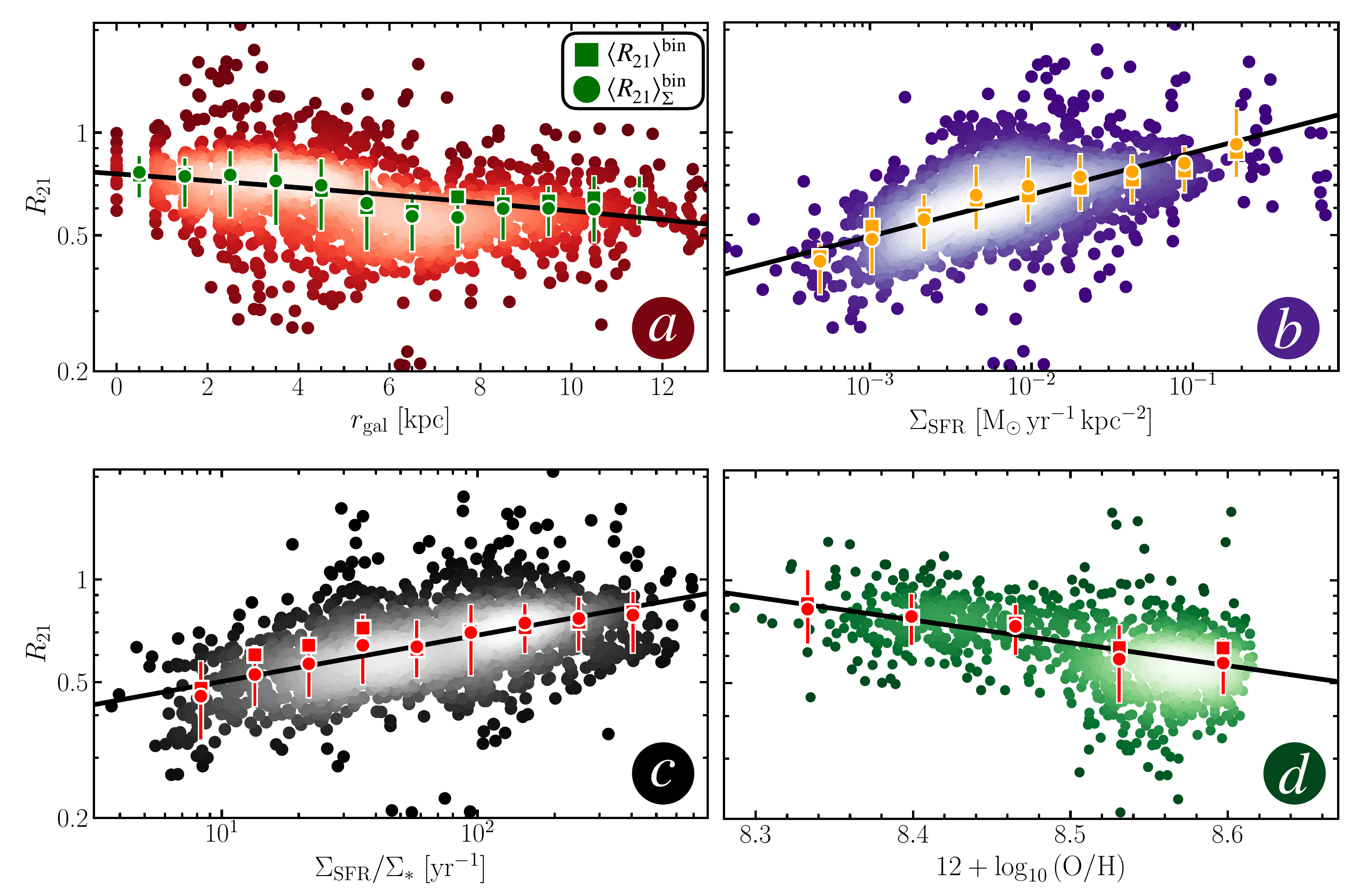}
    \caption{{\bf $R_{21}$ correlation trends combining all sightlines} The panels present the trends for the significantly detected sightlines. The points are color-coded using a density distribution function to illustrate the where the majority of points lie within the respective parameter space. The larger markers with the error bars represent the binned means (circle: equally weighted; square: ratio-of-sums average). The error bars present the 16$^{\rm th}$-to-84$^{\rm th}$ percentile range for individual significant measurements in that bin. (\textit{a}) The line ratio as function of galactocentric radius in kpc. 
    %(\textit{Top Right}) The line ratio normalized by the galaxy-wide mean ratio as function of galactocentric radius normalized by the isophotal radius ($r_{25}$). 
    (\textit{b}) The line ratio as function of SFR surface density, $\Sigma_{\rm SFR}$. 
    %(\textit{Second Row Right}) The line ratio normalized by the galaxy-wide mean ratio as function of $\Sigma_{\rm SFR}$ normalized by the galaxy average. 
    (\textit{c}) The line ratio as function of the local specific SFR ($\Sigma_{\rm SFR}/\Sigma_{\rm *}$). 
    %(\textit{Third Row Right}) The line ratio normalized by the galaxy-wide mean ratio as function of the specific SFR normalized by the galaxy average. 
    (\textit{d}) The line ratio as function of the metallicity, only for the 8 (of 14) galaxies with available metallicity maps. 
    %(\textit{Bottom Right}) The line ratio normalized by the galaxy-wide mean ratio as function of the metallicity normalized by the galaxy-wide average. 
    }
    \label{fig:r21_relation_all}
\end{figure*}

In \autoref{fig:r21_galToGal_trends} we illustrate the galaxy-wide median values plotted against key galaxy-integrated properties, such as the stellar mass, the SFR, and the specific SFR. With a Pearson correlation coefficient of $R_{\rm p}{=}-0.5$ ($p$-value of 0.05), we find that the galaxy-wide $\rtwo$ seems to particularly scale with the stellar mass (and to a limited degree to the specific SFR). {Physically, we expect these galaxy-averaged parameters to correlate with the average intensity of the radiation field, which in turn results in a stronger heating of the gas, driving up the line ratio. For instance, lower-mass galaxies tend to have higher sSFR, and hence a higher local $\Sigma_{\rm SFR}$. In addition, given the galaxy mass-metallicity relation \citep{Tremonti2004, Lee2006}, we expect lower mass galaxies to have lower metallicities, resulting in a lower dust-to-gas ratio and hence more intense radiation fields. These effects could explain the observed anticorrelation with stellar mass and correlation with sSFR. }
%\edit1
{ Given the small number of galaxies within our sample, we caution, however, that individual outliers can affect the observed trends}. For reference, we also include the individual galaxy-wide measurements from \citet{Leroy2022}. %\edit1
{We also include their reported scaling relation for SFR/M$_*$ as well as the trends for the three quantities as derived by \citet{Keenan2024b}. Regarding the individual galaxy-integrated values, }it is evident that our sample of measurements shows a lower scatter. For the residual with the fitted line, we find a scatter of 0.05 dex for our ALMA-based observations vs. 0.11 dex for the single-dish sample, despite covering a similar dynamical range in galaxy SFR and stellar mass. This likely reflects the superior flux calibration stability and improved signal-to-noise of ALMA-based CO line intensity measurements. %In particular, the scatter of 0.11\,dex is consistent with the degree of scatter expected for a 20\% flux calibration uncertainty. For the ALMA-based observations, assuming a calibration uncertainty of 5\%, we would expect a scatter of 0.02\,dex. Our finding of a scatter of 0.05\,dex therefore suggest some additional, physical scatter to the ratio, likely related to the type or morphology of the galaxy themselves.

In our sample, we have two sources that host an AGN (NGC\,3627, \citealt{Goulding2009}; and NGC\,7496, \citealt{Schmitt2006}). To see whether the AGN influences the galaxy-wide average, we highlight them in  \autoref{fig:r21_galToGal_trends} using a blue-colored marker. We do not find any clear evidence that the AGN affects the galaxy-wide CO line ratio median, as these points do not deviate in any significant way from the regression line or the sample-wide median. %\edit1
{We note that any impact of the AGN on the molecular gas will be likely be limited to the central region of the galaxy. At our 1.7\,kpc physical resolution, we therefore do not expect to see much impact from AGN because even our central beam will include GMCs that likely are not significantly affected by the AGN.}

%For reference, we also present the galaxy-wide average $\rtwo$ in the SFR-M$_*$ parameter space in \autoref{fig:r21_SFRMstar}. We note that, in particular, the two early-type galaxies in the Green valley (NGC\,1317 and NGC\,4457) show enhanced ratios. Apart from this, the anticorrelation with the stellar mass is evident with the most massive spiral galaxies showing the overall lowest galaxy-wide average line ratios. %To robustly determine the trend of higher $\rtwo$ for late-type galaxies, a larger sample size is required.
For reference, we also present the galaxy-wide average $\rtwo$ in the SFR-M$_*$ parameter space in \autoref{fig:r21_SFRMstar}. Notably, the two early-type galaxies in the Green valley (NGC,1317 and NGC,4457) show elevated ratios (though this is not exclusive, as two main sequence galaxies also exhibit similar values). Additionally, while there is a{n apparent} trend of decreasing ratios with increasing stellar mass %\edit1
{(see also \autoref{fig:r21_galToGal_trends})}, it is important to note exceptions, particularly the presence of galaxies with higher ratios at both low and high stellar masses. This suggests that the relationship between $\rtwo$ and stellar mass is more complex and  requires multi-parameter analysis to fully characterize for a future study.

\subsubsection{Resolved measurement}\label{sec:res_meas}
In \autoref{fig:r21_relation_all}, we plot \rtwo\ as a function of galactocentric radius ($r_{\rm gal}$), SFR surface density ($\Sigma_{\rm SFR}$), the specific SFR surface density ($\Sigma_{\rm SFR}/\Sigma_\star$), and metallicity ($12+\log_{10} {\rm O/H}$; only for 8/14 targets in our sample). In these panels, we also show the binned medians (circles show equally-weighted values and squares show CO(1-0) intensity weighted ones) and the vertical line illustrates the (equally weighted) 16$^{\rm th}$-to-84$^{\rm th}$ percentile range per bin. 
%We quantify the correlation of all sightlines using the Kendall rank correlation coefficient, $\tau_{\rm Kendall}$. In each panel, we list the Kendall $\tau$'s $p$-value. This correlation coefficient measures the ordinal association between two parameters. In contrast to the Pearson's correlation coefficient, it does not assume a linear correlation between the two quantities. 
In addition, we fit a regression line to the equally weighted binned means. For the regression line, we either assume a linear relation (galactocentric radius) or a power-law\footnote{{We note that the correlations reported here only hold for the dynamical range of the observables. Beyond the covered range, we expect the ratio to behave differently, such as flattening towards unity and stabilizing around values of 0.3 to 0.5 \citep[e.g., ][]{Leroy2022}. See Appendix \ref{app:fitting_presc} for a comparison of different fitting prescriptions.}} (SFR and specific SFR surface density).
We provide an overview of the regression line coefficients in \autoref{table:fitting}.

\begin{deluxetable}{c c c c c }
\tablewidth{0pt}
\tablecaption{Line regression results to binned trends in \autoref{fig:r21_relation_all}. \label{table:fitting}}
\tablehead{\colhead{Average (y)}&\colhead{Quantity (x)} & \colhead{m} &\colhead{q}&\colhead{$\tau_{\rm Kendall}$}
}
\startdata
$\langle R_{21}\rangle$ &\multirow{ 2}{*}{$r_{\rm gal}$} &$-0.017^{+0.006}_{-0.01}$ & $0.76^{+0.08}_{-0.09}$ & $-0.58$ \\
$\langle R_{21}\rangle_{\Sigma}$ & & $-0.012^{+0.004}_{-0.004}$ & $0.74^{+0.08}_{-0.09}$ & $-0.45$  \\\hline
$\log_{10}\langle R_{21}\rangle$ &\multirow{ 2}{*}{$\log_{\rm 10}\left(\Sigma_{\rm SFR}\right)$} &$0.12^{+0.03}_{-0.04}$ & $0.06^{+0.07}_{-0.09}$ & $1.0$\\
$\log_{10}\langle R_{21}\rangle_{\Sigma}$ &&$0.10^{+0.02}_{+0.02}$ & $0.01^{+0.05}_{-0.08}$ & $1.0$  \\\hline
$\log_{10}\langle R_{21}\rangle$ &\multirow{ 2}{*}{$\log_{\rm 10}\left(\Sigma_{\rm SFR}/\Sigma_{\rm *}\right)$}& $0.14^{+0.03}_{-0.03}$ & $-0.43^{+0.09}_{-0.13}$ &  $0.94$ \\
$\log_{10}\langle R_{21}\rangle_{\Sigma}$ & & $0.10^{+0.03}_{-0.03}$ &  $-0.35^{+0.08}_{-0.10}$ & $0.83$  \\\hline
$\langle R_{21}\rangle$ &\multirow{ 2}{*}{$12+\log_{10}\left(\rm O/H\right)$} & $-1.1^{+0.8}_{-0.6}$ &  $9^{+5}_{-8}$ & $-1.0$ \\
$\langle R_{21}\rangle_{\Sigma}$ && $-0.8^{+0.2}_{-0.2}$ &  $8^{+6}_{-6}$ & $-1.0$  \\\hline
\enddata
\tablecomments{We fit a linear regression of the form \mbox{$y = mx+q$}. We also provide the Kendall's $\tau$ correlation coefficient derived for the binned values.}
\end{deluxetable}

\begin{figure*}
    \centering
    \includegraphics[width=\textwidth]{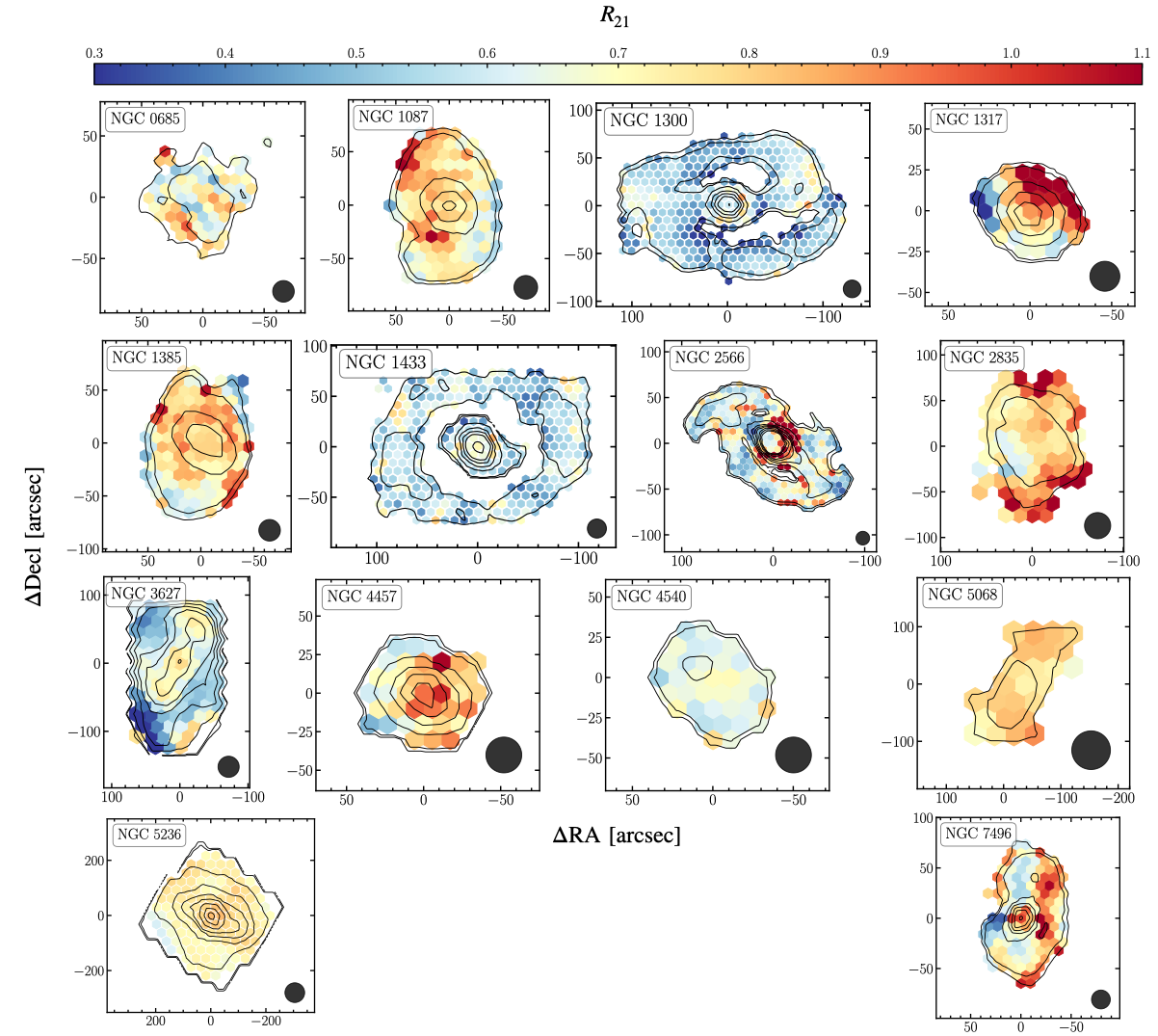}
    \caption{{\bf Resolved $R_{21}$ variation} All maps have been convolved to a common physical resolution of $1.7$\,kpc. Only sightlines for which both the CO(1-0) and CO(2-1) emission are significantly detected at $\rm S/N{>}5$ are shown. The black circle in the lower right corner of each panel presents the beam.  For reference, we illustrate the CO(1-0) integrated intensities using the black contour lines. These are drawn at 0.5, 1, 5, 10, 15, 20, 30, 40, and 50\,K\,km\,s$^{-1}$. }
    \label{fig:r21_maps}
\end{figure*}

\paragraph{Galactocentric radius}
The panel $a$ of \autoref{fig:r21_relation_all} present the correlation of the line ratio with galactocentric radius. We find a negative correlation, as indicated by the negative Kendall $\tau$-value of $-0.58$ or $-0.48$ for the resolved \rtwo\ ratio median and the ratio of sums average respectively (see \autoref{table:fitting}). This trend is consistent with findings from high-resolution (${\sim}200$ pc resolution) observations in NGC\,3627 and NGC\,2903, where the radial variation is particularly driven by the increase of \rtwo\ towards the center \citep{denBrok2023b}. We note that -- qualitatively -- from 6\,kpc onward, $\rtwo$ sees to rise again  in contrast to the trend from 0--6\,kpc. This is likely driven by NGC\,1300, NGC\,1433, and NGC\,2566, which are the galaxies in our sample for which we have radial measurements beyond $r{>}6$\,kpc.

\paragraph{SFR surface density}
Panel $b$ of \autoref{fig:r21_relation_all} illustrates the correlation of $\rtwo$ with the SFR surface density. We measure a significant positive correlation between the two quantities. %, as indicated by a Kendall's correlation rank coefficient's $p$-value close to zero. 
We fit a power-law regression to the binned medians and report a positive correlation with a slope of $m{=}0.12^{+0.03}_{-0.04}$ or $m{=}0.10^{+0.02}_{-0.02}$ respectively to the different binning techniques.

\begin{figure*}
    \centering
    \includegraphics[width=0.95\textwidth]{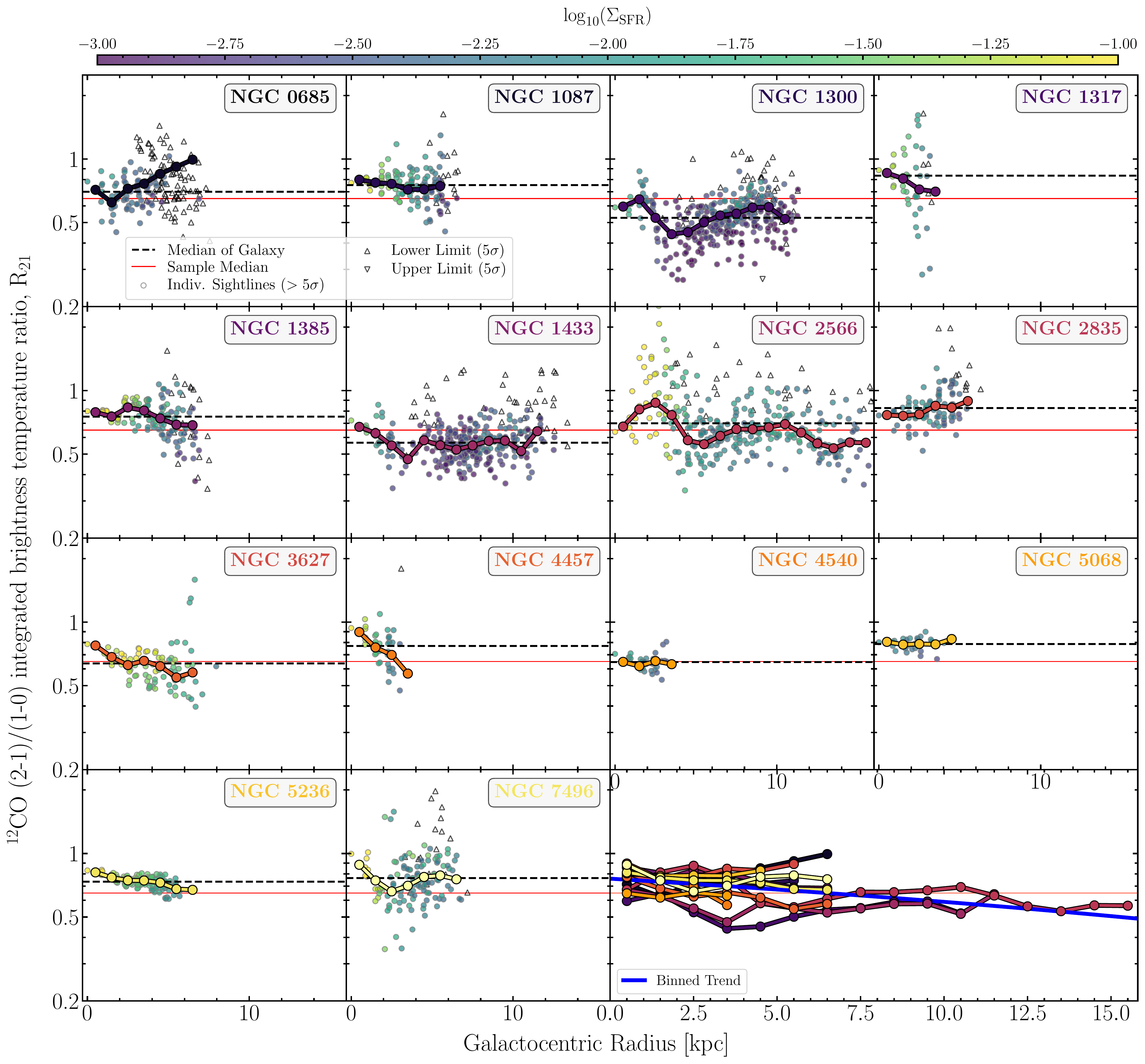}
    \caption{{\bf Radial \rtwo\ variation} The colored points represent significantly detected ($\rm S/N{>}5$) individual sightlines with adjacent lines of sight spaced by one half the beam width. Data with lower signal to noise are plotted using triangles either as upper or lower limits, if only one of the two CO transitions is significantly detected. The galaxy-wide median value for each individual galaxy is plotted as a dashed horizontal line. For reference, we plot all the individual binned line trends in the lower right panel. The red dashed line represents the sample-wide CO line ratio median.  }
    \label{fig:r21_radial_trends}
\end{figure*}

\paragraph{Local specific SFR}

The bottom right panel ($c$) of \autoref{fig:r21_relation_all} presents the correlation with the specific SFR along the sightline. For the correlation with the absolute line ratio, we find again a strong correlation with a Kendall's $\tau$ equal to 1 for the binned trend (see \autoref{table:fitting}). The power-law regression {for SFR surface densities in the $10^{-3} -10^{-1} \rm M_{\odot}\,yr^{-1}\,kpc^{-1}$ range}  yields a slope of $m{=}0.14^{+0.03}_{-0.03}$ for the binned $\rtwo$ median. %When we normalize, the slope is even steeper with $m=0.16$.

\paragraph{Metallicity}
We only have metallicity maps for 8 of 14 galaxies in our sample. In panel $d$ we therefore only show sightlines for those 8 galaxies. We also note that overall, we only cover a small dynamical range of metallicities (${\sim}$0.25\,dex), therefore, any real trends can potentially be hidden in the scatter of our measurements. For $\rtwo$, we measure a significant negative correlation with the metallicity with a Kendall's $\tau$ correlation coefficient of $-1$ (see \autoref{table:fitting}). 
The linear regression through the binned medians yields a negative slope of $m=-1.1^{+0.8}_{-0.6}$.

This negative relation between CO line ratios and metallicity has been reported before \citep[e.g.,][]{Lequeux1994,Kepley2016,Cicone2017} and can be linked generally to the radial trend (c.f., radial metallicity trends) and, in the broader context, to the negative correlation we find with the galaxy integrated M$_*$ (see \autoref{fig:r21_galToGal_trends}) given the mass–metallicity relation \citep[e.g.][]{Tremonti2004}. At lower metallicities, the environment will also exhibit lower dust-to-gas ratios \citep{Leroy2011,Casasola2020} and hence less shielding and harder radiation fields probably driving both, more abundant and stronger excitation.

\subsection{Variation of \texorpdfstring{$R_{21}$}{Lg} within individual galaxies}
We also investigate how the line ratio varies within the individual galaxies of our sample. \autoref{fig:r21_maps} illustrates the resolved $\rtwo$ measurements per galaxy at 1.7 kpc physical resolution. From a qualitative inspection alone, it is evident that the line ratio clearly varies within the galaxies. For instance, 5/14 sources show a clear increase of the line ratio towards the center (e.g., NGC1300, NGC1433, NGC3627, NGC4457, NGC7496) upon visual inspection.

\begin{figure*}
    \centering
    \includegraphics[width=0.95\textwidth]{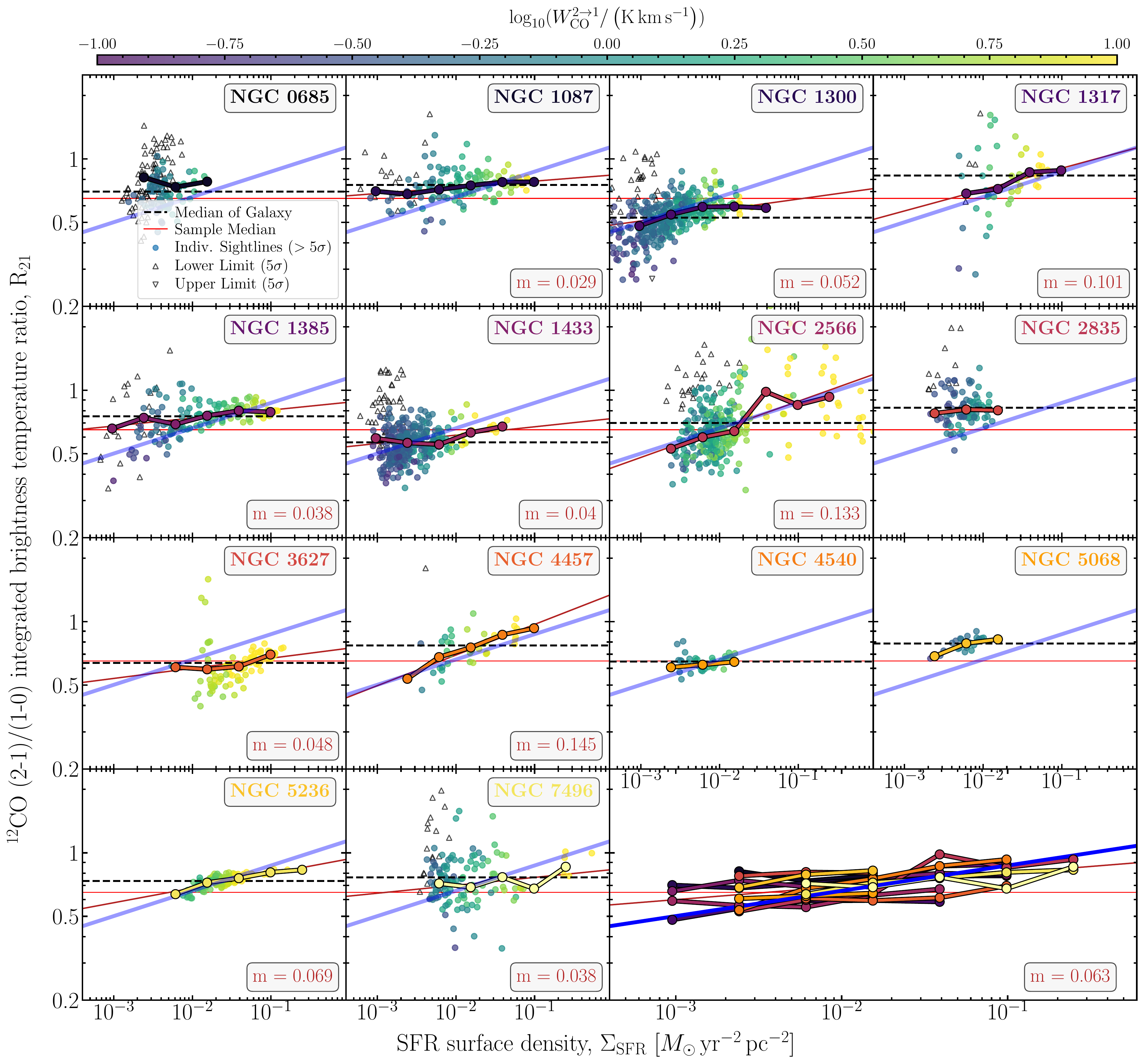}
    \caption{{\bf $R_{21}$ variation with SFR surface density} This figure follows in layout and design \autoref{fig:r21_radial_trends}. Here, the significantly detected sightlines are colored by their corresponding CO(2-1) integrated brightness temperature. In case more than four stacked points are available, we also report the slope of the power-law regression fitted to the stacked trend.  }
    \label{fig:r21_sfr_trends}
\end{figure*}

\paragraph{Galactocentric radius}

In \autoref{fig:r21_radial_trends}, we present the radial trends of the line ratio for each galaxy in our sample. Each panel also presents the stacked line ratio trend by stacking together all sightlines per bin irrespective of their S/N. For reference, we also plot the stacked line trends all together in a single panel in the lower right. Here we use the spectral stacking technique to quantify the radial trends. We show only stacked points where both stacked CO lines are significantly detected ($\rm S/N{>}5$).

Looking at the stacked trends, we note that 5/14 galaxies show radial variations in terms of difference of the median line ratio for center and disk, which we report in \autoref{table:results_ratio}. NGC1300, NGC1433, NGC2566, and NGC7496 show a clear increase towards the center with a flat trend of $\rtwo$ across the disk. NGC3627, in contrast, shows a continuous decreasing trend of $\rtwo$ with galactocentric radius.
Overall, the centers show a range of variation, which argues against a simple picture of more radiation leading to stronger excitation in centers. Likely, the presence of bars and the degree of turbulence play a critical role. To investigate this, however, higher angular resolution are necessary that resolve scales where we can study the gas dynamics and kinematics.
At larger radii, we include a larger amount of low S/N data in our stack. Consequently, we are cautious about drawing overly definitive conclusions. Obtaining sensitive multi-line observations of outer discs will provide valuable insights into whether there is a substantial change in CO excitation in the outer regions of disc galaxies. 

\begin{figure*}
    \centering
    \includegraphics[width=0.95\textwidth]{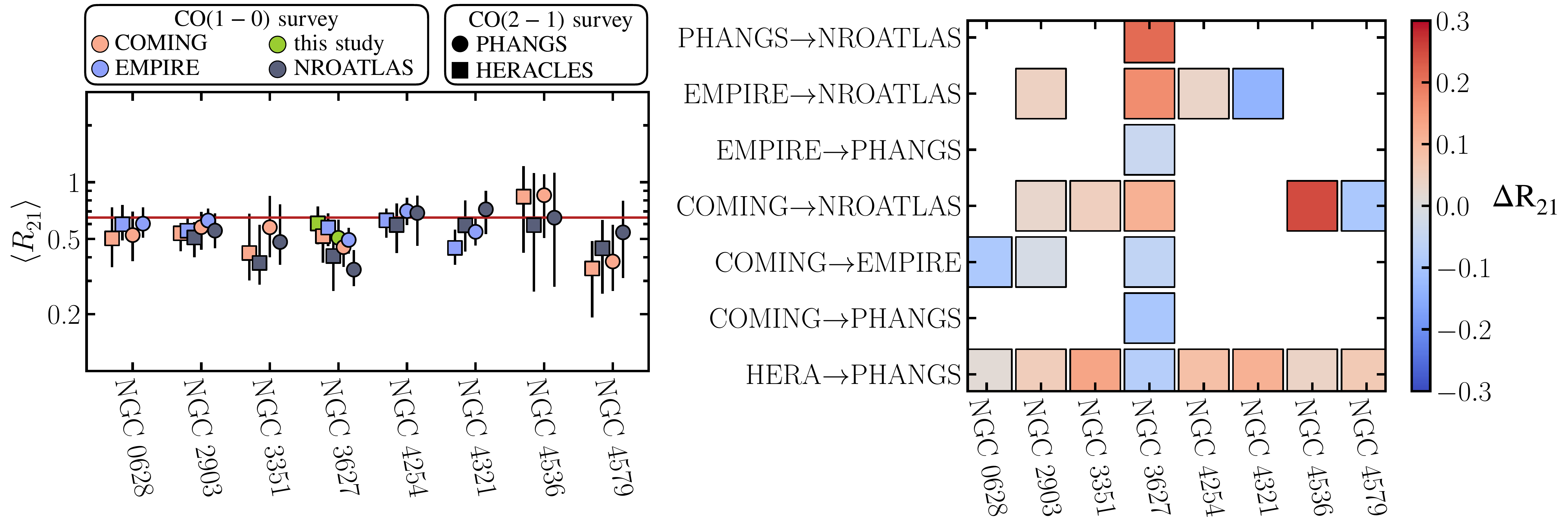}
    \caption{{\bf Estimating instrumental-dependent scatter in $R_{21}$}. For a number of galaxies, we have CO(1-0) and/or CO(2-1) maps from different surveys, enabling up to 8 permutations of CO line ratios (e.g., for NGC\,3627). To estimate the scatter, we compare the $R_{21}$ distribution using different permutations of CO(1-0) and CO(2-1) to compute the line ratio. (\textit{Left}) The (volume-weighted) mean and 16$^{\rm th}$-to-84$^{\rm th}$ percentile range. Square markers indicate CO(2-1) from HERACLES \citep{Leroy2009} and circle markers indicate CO(2-1) from PHANGS \citep{Leroy2021} was used. The color of the marker indicates which CO(1-0) dataset is used. The horizontal red line illustrates the galaxy-wide $R_{21}$ average derived in this study. (\textit{Right}) The absolute change in the average $R_{21}$ value based on substituting either the CO(1-0) or CO(2-1) data by a different survey, as indicated on the $y$-axis.  }
    \label{fig:comp_literature_permut}
\end{figure*}

\paragraph{SFR surface density}
In \autoref{fig:r21_sfr_trends}, we investigate the correlation of the CO line ratio with the SFR surface density per galaxy. Again, we use spectral stacking to quantify the overall trends. Previous studies suggest a relation of the CO ratio with SFR surface density. This is because higher SFR surface densities trace regions of  higher-density gas \citep{Usero2015, Gallagher2018}, stronger radiation fields due to young stars present shortly after star formation \citep{Narayanan2014}, and higher cosmic-ray densities.
In our previous analysis when looking at all sightlines combined, we measured a positive correlation, with a slope of ${\sim}0.1$ for the binned medians (see \autoref{fig:r21_relation_all}). For galaxies for which we have more than four independently stacked CO line ratio measurements, we repeat the measurement and perform a linear regression in the form of a power law to the stacked data points. We report the resulting slope in each panel.  

We find a large range in slopes for the CO line ratio trends with SFR surface density. Some of the galaxies in our sample do not show any or only a mild correlation between the two quantities. For instance, NGC1087, NGC1300, NGC1385, NGC1433, and NGC 7496 all show slopes that are less than 0.05.  When we perform a liner regression when combining all the stacked trends, we get a slope of $m=0.06$, which is shallower than the trend found when looking at the individual lines of sight. Overall, however, our results confirm that internal correlations exist between star formation activity and CO excitation in nearby, normal, star-forming galaxies. Nevertheless, to provide a more comprehensive conclusion a larger sample of the galaxy population and at higher angular resolutions (to ensure resolving individual environments) will be necessary to explore potential variations across different environments.

\begin{figure*}
    \centering
    \vspace{5mm}\includegraphics[width=\textwidth]{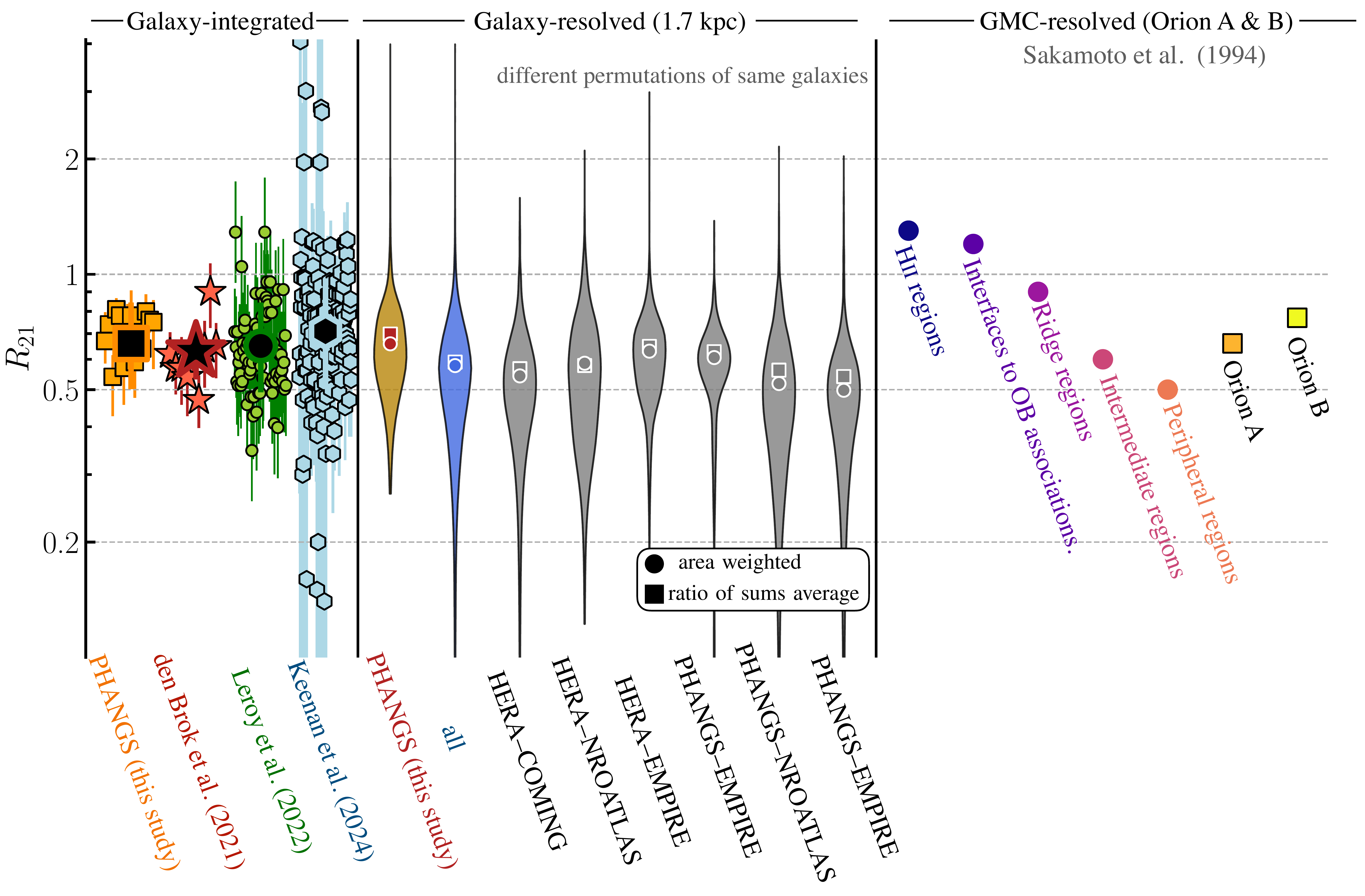}
    \caption{{\bf Comparing $R_{21}$ ratio surveys at different scales}. %\edit1
    {(Galaxy integrated) Each marker represents the median line ratio per galaxy. We also illustrate the sample wide median and 16$^{\rm th}$-to-$84^{\rm th}$ percentile range using a larger, grey-colored marker.  The $y$ axis carries no information. The markers are slightly offset along the $y$ axis, such that points are individually recognizable. (Galaxy resolved) Histogram of sightline $R_{21}$ value distribution for different permutations of surveys. The volume and intensity weighted averages are represented with a circle or square respectively for each distribution. The dark orange distribution is using the data presented in this study. The blue distribution is combining all data from all surveys. The grey distributions represent the results using a different survey for CO(2-1) (first name in $x$-axis) and CO(1-0) (second name). The dotted line represents the average value derived by this study. (GMC-resolved) For reference, the panel also presents the reported average value for GMC-resolved observations using the Orion A and B clouds as reference. The values are adopted from \citet{Sakamoto1994}.}}
    \label{fig:comp_samples}
\end{figure*}

\begin{figure}
    \centering
    \includegraphics[width=0.95\columnwidth]{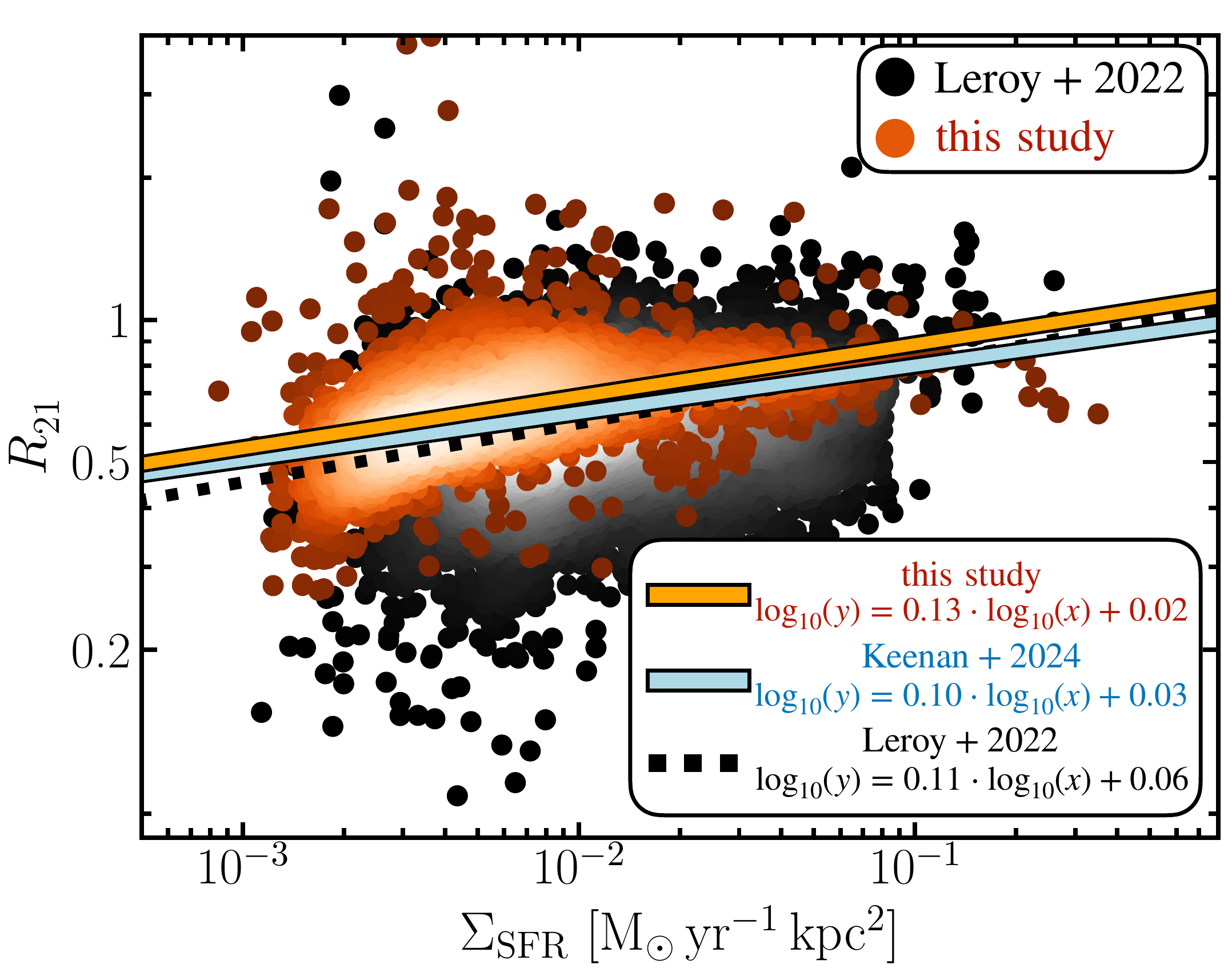}
    \caption{{\bf $R_{21}$ literature sightline comparison}.  
    The points show the correlation of the line ratio with the SFR surface density. The dark orange points indicate the data from this study, the black points from \citet{Leroy2022}. The red solid line is a linear regression fit (in log-log space) through the individually detected sightlines. The blue line presents the fit reported by \citet{Keenan2024a} and the black dotted line represents the fit derived by \citet{Leroy2022}. }
    \label{fig:comp_literature_r21}
\end{figure}

{\renewcommand{\arraystretch}{0.9}
\begin{deluxetable*}{l c c c c c c }
\tablewidth{0pt} 
\tablecaption{ 
Overview of which CO(1-0) and CO(2-1) mapping surveys cover which galaxies \edit1{from all available surveys}.\label{tb:mapping_surveys}}
\tablehead{
&\multicolumn{4}{c}{CO(1-0)}&\multicolumn{2}{c}{CO(2-1)} \\
\colhead{Galaxy} & \colhead{ALMA (cycle 9)} & \colhead{EMPIRE}& \colhead{COMING}& \colhead{NROATLAS}&\colhead{HERACLES}&\colhead{PHANGS}
}
\startdata
%NGC 0253&\textcolor{red}{\ding{56}}&\textcolor{red}{\ding{56}}&\textcolor{red}{\ding{56}}&\textcolor{green}{\ding{52}}&\textcolor{red}{\ding{56}}&\textcolor{green}{\ding{52}} \\
%NGC 0337&\textcolor{red}{\ding{56}}&\textcolor{red}{\ding{56}}&\textcolor{green}{\ding{52}}&\textcolor{red}{\ding{56}}&\textcolor{green}{\ding{52}}&\textcolor{red}{\ding{56}} \\
%NGC 0628 &\textcolor{red}{\ding{56}}&\textcolor{green}{\ding{52}}&\textcolor{green}{\ding{52}}&\textcolor{red}{\ding{56}}&\textcolor{green}{\ding{52}}&\textcolor{green}{\ding{52}} \\
%&\multicolumn{6}{c}{$\vdots$}\\
%NGC 7331 &\textcolor{red}{\ding{56}}&\textcolor{red}{\ding{56}}&\textcolor{green}{\ding{52}}&\textcolor{red}{\ding{56}}&\textcolor{green}{\ding{52}}&\textcolor{red}{\ding{56}} \\
%NGC 7496 &\textcolor{green}{\ding{52}}&\textcolor{red}{\ding{56}}&\textcolor{red}{\ding{56}}&\textcolor{red}{\ding{56}}&\textcolor{red}{\ding{56}}&\textcolor{green}{\ding{52}} \\
\scriptsize NGC 0253 & \textcolor{red}{\ding{56}} &\textcolor{red}{\ding{56}} &\textcolor{red}{\ding{56}} &\textcolor{green}{\ding{52}} &\textcolor{red}{\ding{56}} &\textcolor{green}{\ding{52}} \\
\scriptsize NGC 0337 & \textcolor{red}{\ding{56}} &\textcolor{red}{\ding{56}} &\textcolor{green}{\ding{52}} &\textcolor{red}{\ding{56}} &\textcolor{green}{\ding{52}} &\textcolor{red}{\ding{56}} \\
\scriptsize NGC 0628 & \textcolor{red}{\ding{56}} &\textcolor{green}{\ding{52}} &\textcolor{green}{\ding{52}} &\textcolor{red}{\ding{56}} &\textcolor{green}{\ding{52}} &\textcolor{green}{\ding{52}} \\
\scriptsize NGC 0685 & \textcolor{green}{\ding{52}} &\textcolor{red}{\ding{56}} &\textcolor{red}{\ding{56}} &\textcolor{red}{\ding{56}} &\textcolor{red}{\ding{56}} &\textcolor{green}{\ding{52}} \\
\scriptsize NGC 1068 & \textcolor{red}{\ding{56}} &\textcolor{red}{\ding{56}} &\textcolor{red}{\ding{56}} &\textcolor{green}{\ding{52}} &\textcolor{red}{\ding{56}} &\textcolor{green}{\ding{52}} \\
\scriptsize NGC 1087 & \textcolor{green}{\ding{52}} &\textcolor{red}{\ding{56}} &\textcolor{green}{\ding{52}} &\textcolor{red}{\ding{56}} &\textcolor{red}{\ding{56}} &\textcolor{green}{\ding{52}} \\
\scriptsize NGC 1300 & \textcolor{green}{\ding{52}} &\textcolor{red}{\ding{56}} &\textcolor{red}{\ding{56}} &\textcolor{red}{\ding{56}} &\textcolor{red}{\ding{56}} &\textcolor{green}{\ding{52}} \\
\scriptsize NGC 1317 & \textcolor{green}{\ding{52}} &\textcolor{red}{\ding{56}} &\textcolor{red}{\ding{56}} &\textcolor{red}{\ding{56}} &\textcolor{red}{\ding{56}} &\textcolor{green}{\ding{52}} \\
\scriptsize NGC 1385 & \textcolor{green}{\ding{52}} &\textcolor{red}{\ding{56}} &\textcolor{red}{\ding{56}} &\textcolor{red}{\ding{56}} &\textcolor{red}{\ding{56}} &\textcolor{green}{\ding{52}} \\
\scriptsize NGC 1433 & \textcolor{green}{\ding{52}} &\textcolor{red}{\ding{56}} &\textcolor{red}{\ding{56}} &\textcolor{red}{\ding{56}} &\textcolor{red}{\ding{56}} &\textcolor{green}{\ding{52}} \\
\scriptsize NGC 2146 & \textcolor{red}{\ding{56}} &\textcolor{red}{\ding{56}} &\textcolor{green}{\ding{52}} &\textcolor{red}{\ding{56}} &\textcolor{green}{\ding{52}} &\textcolor{red}{\ding{56}} \\
\scriptsize NGC 2566 & \textcolor{green}{\ding{52}} &\textcolor{red}{\ding{56}} &\textcolor{red}{\ding{56}} &\textcolor{red}{\ding{56}} &\textcolor{red}{\ding{56}} &\textcolor{green}{\ding{52}} \\
\scriptsize NGC 2775 & \textcolor{red}{\ding{56}} &\textcolor{red}{\ding{56}} &\textcolor{green}{\ding{52}} &\textcolor{red}{\ding{56}} &\textcolor{red}{\ding{56}} &\textcolor{green}{\ding{52}} \\
\scriptsize NGC 2835 & \textcolor{green}{\ding{52}} &\textcolor{red}{\ding{56}} &\textcolor{red}{\ding{56}} &\textcolor{red}{\ding{56}} &\textcolor{red}{\ding{56}} &\textcolor{green}{\ding{52}} \\
\scriptsize NGC 2841 & \textcolor{red}{\ding{56}} &\textcolor{red}{\ding{56}} &\textcolor{green}{\ding{52}} &\textcolor{red}{\ding{56}} &\textcolor{green}{\ding{52}} &\textcolor{red}{\ding{56}} \\
\scriptsize NGC 2903 & \textcolor{red}{\ding{56}} &\textcolor{green}{\ding{52}} &\textcolor{green}{\ding{52}} &\textcolor{green}{\ding{52}} &\textcolor{green}{\ding{52}} &\textcolor{green}{\ding{52}} \\
\scriptsize NGC 2976 & \textcolor{red}{\ding{56}} &\textcolor{red}{\ding{56}} &\textcolor{green}{\ding{52}} &\textcolor{red}{\ding{56}} &\textcolor{green}{\ding{52}} &\textcolor{red}{\ding{56}} \\
\scriptsize NGC 3077 & \textcolor{red}{\ding{56}} &\textcolor{red}{\ding{56}} &\textcolor{green}{\ding{52}} &\textcolor{red}{\ding{56}} &\textcolor{green}{\ding{52}} &\textcolor{red}{\ding{56}} \\
\scriptsize NGC 3147 & \textcolor{red}{\ding{56}} &\textcolor{red}{\ding{56}} &\textcolor{green}{\ding{52}} &\textcolor{red}{\ding{56}} &\textcolor{green}{\ding{52}} &\textcolor{red}{\ding{56}} \\
\scriptsize NGC 3184 & \textcolor{red}{\ding{56}} &\textcolor{green}{\ding{52}} &\textcolor{red}{\ding{56}} &\textcolor{green}{\ding{52}} &\textcolor{green}{\ding{52}} &\textcolor{red}{\ding{56}} \\
\scriptsize NGC 3198 & \textcolor{red}{\ding{56}} &\textcolor{red}{\ding{56}} &\textcolor{green}{\ding{52}} &\textcolor{red}{\ding{56}} &\textcolor{green}{\ding{52}} &\textcolor{red}{\ding{56}} \\
\scriptsize NGC 3351 & \textcolor{red}{\ding{56}} &\textcolor{red}{\ding{56}} &\textcolor{green}{\ding{52}} &\textcolor{green}{\ding{52}} &\textcolor{green}{\ding{52}} &\textcolor{green}{\ding{52}} \\
\scriptsize NGC 3521 & \textcolor{red}{\ding{56}} &\textcolor{red}{\ding{56}} &\textcolor{green}{\ding{52}} &\textcolor{red}{\ding{56}} &\textcolor{green}{\ding{52}} &\textcolor{green}{\ding{52}} \\
\scriptsize NGC 3627 & \textcolor{green}{\ding{52}} &\textcolor{green}{\ding{52}} &\textcolor{green}{\ding{52}} &\textcolor{green}{\ding{52}} &\textcolor{green}{\ding{52}} &\textcolor{green}{\ding{52}} \\
\scriptsize NGC 3631 & \textcolor{red}{\ding{56}} &\textcolor{red}{\ding{56}} &\textcolor{red}{\ding{56}} &\textcolor{red}{\ding{56}} &\textcolor{green}{\ding{52}} &\textcolor{red}{\ding{56}} \\
\scriptsize NGC 3938 & \textcolor{red}{\ding{56}} &\textcolor{red}{\ding{56}} &\textcolor{green}{\ding{52}} &\textcolor{red}{\ding{56}} &\textcolor{green}{\ding{52}} &\textcolor{red}{\ding{56}} \\
\scriptsize NGC 4030 & \textcolor{red}{\ding{56}} &\textcolor{red}{\ding{56}} &\textcolor{green}{\ding{52}} &\textcolor{red}{\ding{56}} &\textcolor{green}{\ding{52}} &\textcolor{red}{\ding{56}} \\
\scriptsize NGC 4051 & \textcolor{red}{\ding{56}} &\textcolor{red}{\ding{56}} &\textcolor{red}{\ding{56}} &\textcolor{red}{\ding{56}} &\textcolor{green}{\ding{52}} &\textcolor{red}{\ding{56}} \\
\scriptsize NGC 4214 & \textcolor{red}{\ding{56}} &\textcolor{red}{\ding{56}} &\textcolor{green}{\ding{52}} &\textcolor{red}{\ding{56}} &\textcolor{green}{\ding{52}} &\textcolor{red}{\ding{56}} \\
\scriptsize NGC 4254 & \textcolor{red}{\ding{56}} &\textcolor{green}{\ding{52}} &\textcolor{red}{\ding{56}} &\textcolor{green}{\ding{52}} &\textcolor{green}{\ding{52}} &\textcolor{green}{\ding{52}} \\
\scriptsize NGC 4303 & \textcolor{red}{\ding{56}} &\textcolor{red}{\ding{56}} &\textcolor{green}{\ding{52}} &\textcolor{red}{\ding{56}} &\textcolor{green}{\ding{52}} &\textcolor{green}{\ding{52}} \\
\scriptsize NGC 4321 & \textcolor{red}{\ding{56}} &\textcolor{green}{\ding{52}} &\textcolor{red}{\ding{56}} &\textcolor{green}{\ding{52}} &\textcolor{green}{\ding{52}} &\textcolor{green}{\ding{52}} \\
\scriptsize NGC 4457 & \textcolor{green}{\ding{52}} &\textcolor{red}{\ding{56}} &\textcolor{red}{\ding{56}} &\textcolor{red}{\ding{56}} &\textcolor{red}{\ding{56}} &\textcolor{green}{\ding{52}} \\
\scriptsize NGC 4535 & \textcolor{red}{\ding{56}} &\textcolor{red}{\ding{56}} &\textcolor{red}{\ding{56}} &\textcolor{red}{\ding{56}} &\textcolor{green}{\ding{52}} &\textcolor{green}{\ding{52}} \\
\scriptsize NGC 4536 & \textcolor{red}{\ding{56}} &\textcolor{red}{\ding{56}} &\textcolor{green}{\ding{52}} &\textcolor{green}{\ding{52}} &\textcolor{green}{\ding{52}} &\textcolor{green}{\ding{52}} \\
\scriptsize NGC 4540 & \textcolor{green}{\ding{52}} &\textcolor{red}{\ding{56}} &\textcolor{red}{\ding{56}} &\textcolor{red}{\ding{56}} &\textcolor{red}{\ding{56}} &\textcolor{green}{\ding{52}} \\
\scriptsize NGC 4548 & \textcolor{red}{\ding{56}} &\textcolor{red}{\ding{56}} &\textcolor{red}{\ding{56}} &\textcolor{red}{\ding{56}} &\textcolor{green}{\ding{52}} &\textcolor{green}{\ding{52}} \\
\scriptsize NGC 4569 & \textcolor{red}{\ding{56}} &\textcolor{red}{\ding{56}} &\textcolor{red}{\ding{56}} &\textcolor{green}{\ding{52}} &\textcolor{green}{\ding{52}} &\textcolor{green}{\ding{52}} \\
\scriptsize NGC 4579 & \textcolor{red}{\ding{56}} &\textcolor{red}{\ding{56}} &\textcolor{green}{\ding{52}} &\textcolor{green}{\ding{52}} &\textcolor{green}{\ding{52}} &\textcolor{green}{\ding{52}} \\
\scriptsize NGC 4654 & \textcolor{red}{\ding{56}} &\textcolor{red}{\ding{56}} &\textcolor{red}{\ding{56}} &\textcolor{red}{\ding{56}} &\textcolor{green}{\ding{52}} &\textcolor{green}{\ding{52}} \\
\scriptsize NGC 4666 & \textcolor{red}{\ding{56}} &\textcolor{red}{\ding{56}} &\textcolor{green}{\ding{52}} &\textcolor{red}{\ding{56}} &\textcolor{green}{\ding{52}} &\textcolor{red}{\ding{56}} \\
\scriptsize NGC 4689 & \textcolor{red}{\ding{56}} &\textcolor{red}{\ding{56}} &\textcolor{red}{\ding{56}} &\textcolor{red}{\ding{56}} &\textcolor{green}{\ding{52}} &\textcolor{green}{\ding{52}} \\
\scriptsize NGC 4736 & \textcolor{red}{\ding{56}} &\textcolor{red}{\ding{56}} &\textcolor{red}{\ding{56}} &\textcolor{green}{\ding{52}} &\textcolor{green}{\ding{52}} &\textcolor{red}{\ding{56}} \\
\scriptsize NGC 5055 & \textcolor{red}{\ding{56}} &\textcolor{green}{\ding{52}} &\textcolor{green}{\ding{52}} &\textcolor{green}{\ding{52}} &\textcolor{green}{\ding{52}} &\textcolor{red}{\ding{56}} \\
\scriptsize NGC 5068 & \textcolor{green}{\ding{52}} &\textcolor{red}{\ding{56}} &\textcolor{red}{\ding{56}} &\textcolor{red}{\ding{56}} &\textcolor{red}{\ding{56}} &\textcolor{green}{\ding{52}} \\
\scriptsize NGC 5194 & \textcolor{red}{\ding{56}} &\textcolor{green}{\ding{52}} &\textcolor{red}{\ding{56}} &\textcolor{red}{\ding{56}} &\textcolor{green}{\ding{52}} &\textcolor{red}{\ding{56}} \\
\scriptsize NGC 5236 & \textcolor{green}{\ding{52}} &\textcolor{red}{\ding{56}} &\textcolor{red}{\ding{56}} &\textcolor{red}{\ding{56}} &\textcolor{red}{\ding{56}} &\textcolor{green}{\ding{52}} \\
\scriptsize NGC 5248 & \textcolor{red}{\ding{56}} &\textcolor{red}{\ding{56}} &\textcolor{green}{\ding{52}} &\textcolor{red}{\ding{56}} &\textcolor{green}{\ding{52}} &\textcolor{green}{\ding{52}} \\
\scriptsize NGC 5364 & \textcolor{red}{\ding{56}} &\textcolor{red}{\ding{56}} &\textcolor{green}{\ding{52}} &\textcolor{red}{\ding{56}} &\textcolor{green}{\ding{52}} &\textcolor{red}{\ding{56}} \\
\scriptsize NGC 5457 & \textcolor{red}{\ding{56}} &\textcolor{red}{\ding{56}} &\textcolor{red}{\ding{56}} &\textcolor{green}{\ding{52}} &\textcolor{green}{\ding{52}} &\textcolor{red}{\ding{56}} \\
\scriptsize NGC 5713 & \textcolor{red}{\ding{56}} &\textcolor{red}{\ding{56}} &\textcolor{green}{\ding{52}} &\textcolor{red}{\ding{56}} &\textcolor{green}{\ding{52}} &\textcolor{red}{\ding{56}} \\
\scriptsize NGC 6946 & \textcolor{red}{\ding{56}} &\textcolor{green}{\ding{52}} &\textcolor{red}{\ding{56}} &\textcolor{green}{\ding{52}} &\textcolor{green}{\ding{52}} &\textcolor{red}{\ding{56}} \\
\scriptsize NGC 7331 & \textcolor{red}{\ding{56}} &\textcolor{red}{\ding{56}} &\textcolor{green}{\ding{52}} &\textcolor{red}{\ding{56}} &\textcolor{green}{\ding{52}} &\textcolor{red}{\ding{56}} \\
\scriptsize NGC 7496 & \textcolor{green}{\ding{52}} &\textcolor{red}{\ding{56}} &\textcolor{red}{\ding{56}} &\textcolor{red}{\ding{56}} &\textcolor{red}{\ding{56}} &\textcolor{green}{\ding{52}} \\
\enddata
%\tablecomments{This table is available in machine-readable format in its entirety. }
\end{deluxetable*}
}

\subsection{Scatter of the line ratio}
For a number of galaxies, we have CO(1-0) and/or CO(2-1) observations from different mapping surveys. Therefore, we can attempt to quantify instrumental-dependent scatter and systematic variation in the $R_{21}$ line ratio value distribution by systematically assessing the different permutations of surveys to compute the line ratio. In \autoref{fig:comp_literature_permut}, we present the distribution of line ratio values per sightline for different combinations of surveys. For eight galaxies, we have CO(1-0) and CO(2-1) mapping coverage from multiple surveys. The left panel in \autoref{fig:comp_literature_permut} shows the area weighted average and 16$^{\rm th}$-to-84$^{\rm th}$ percentile range for the different survey permutations per galaxy. For this comparison, we convolved all different CO(1-0) and CO(2-1) maps to the same resolution and only include sightlines per galaxy where we detect (${>}5\sigma$) CO emission in all surveys. In the right panel of \autoref{fig:comp_literature_permut}, we quantify the change (in terms of absolute difference in mean $\rtwo$) for a given substitution of the survey. For reference, we list the means and 16$^{\rm th}$-to-84$^{\rm th}$ percentile range of the line ratio distributions for the different survey permutations in \autoref{tb:results}. 

From the left panel of \autoref{fig:comp_literature_permut}, it is evident that the average can systematically differ by up to 0.3\,dex for the different permutations. For instance, for NGC\,3627, the equally weighted average ranges from $0.34^{+0.09}_{-0.06}$ when using CO(1-0) from NROATLAS and CO(2-1) from HERACLES to $0.60^{+0.13}_{-0.13}$ when using ALMA CO(1-0) and CO(2-1) observations\footnote{We note, however, that  NGC\,3627 was a HERACLES pilot study. The observations were of lower quality, which can partially explain the large variation with other observations of NGC\,3627.}.  Looking at the right panel in \autoref{fig:comp_literature_permut}, there seem to be systematic changes when substituting different surveys. For instance, when substituting the CO(2-1) observations from the HERACLES survey with the data from the PHANGS survey, the resulting $\rtwo$ is consistently larger by an absolute value of ${\sim}0.1$, apart for NGC\,3627. The overall systematic offsets in the average line ratio of 10\%--20\% are expected within the uncertainty of flux calibration.

We note that this telescope-dependent variation is on a similar magnitude (i.e., 10\%--20\%)  to the overall dynamical range we find for the line ratio. 
We emphasize again that in this analysis, we account for the instrumental variation which does not contribute to the scatter observed within each individual galaxy. We also note that the the line ratio average variation can be to a certain degree result from the different sensitivity of various surveys. In our cross-sample comparison, however, we ensured to use sightlines that have a detection in all surveys, therefore limiting such bias (but we note that this leads to a bias to the more CO luminous sightlines).  The finding of the scatter within each galaxy being greater than the galaxy-to-galaxy variation supports the same conclusion that the scatter within a galaxy (${\sim}0.1$--$0.2$\,dex) is more significant than the galaxy-to-galaxy scatter (${\sim}0.05$\,dex).  Such galaxy-internal variation are mostly driven by different environments, e.g. center, bars, and the contrast between spiral arm and interarm. %At kpc-scale resolutions, these environments can get confused, driving up the point-to-point scatter. 

\begin{deluxetable*}{l c c c c c c c c }
\tablewidth{0pt}
\tablecaption{ 
Summary of volume-weighted average line ratio for different permutations\label{tb:results}}
\tablehead{\colhead{CO(2-1) survey}&\multicolumn{4}{c}{PHANGS}&\multicolumn{4}{c}{HERACLES}
\\
\colhead{CO(1-0) survey}&\colhead{COMING}&\colhead{ALMA}&\colhead{EMPIRE}&\colhead{NROATLAS}&\colhead{COMING}&\colhead{ALMA}&\colhead{EMPIRE}&\colhead{NROATLAS}
}
\startdata
%NGC 0628 & $0.50^{0.22}_{0.14}$& -- & $0.60^{0.14}_{0.09}$& --  & $0.52^{0.16}_{0.14}$& -- & $0.60^{0.12}_{0.09}$& --   \\
NGC\,0628 & $0.50^{+0.22}_{-0.14}$ & -- & $0.60^{+0.15}_{-0.10}$  & -- & $0.52^{+0.16}_{-0.14}$ & -- & $0.60^{+0.12}_{-0.09}$ & --\\
NGC\,2903 & $0.54^{+0.06}_{-0.10}$ & -- & $0.55^{+0.09}_{-0.04}$  & $0.51^{+0.10}_{-0.10}$ & $0.58^{+0.11}_{-0.13}$ & -- & $0.63^{+0.09}_{-0.07}$ & $0.55^{+0.12}_{-0.10}$\\
NGC\,3351 & $0.42^{+0.25}_{-0.12}$ & -- & --  & $0.37^{+0.21}_{-0.08}$ & $0.58^{+0.26}_{-0.17}$ & -- & -- & $0.48^{+0.27}_{-0.11}$\\
NGC\,3627 & $0.52^{+0.09}_{-0.14}$ & $0.60^{+0.13}_{-0.13}$ & $0.57^{+0.10}_{-0.11}$  & $0.41^{+0.11}_{-0.14}$ & $0.45^{+0.05}_{-0.09}$ & $0.51^{+0.12}_{-0.06}$ & $0.49^{+0.07}_{-0.07}$ & $0.34^{+0.09}_{-0.06}$\\
NGC\,4254 & -- & -- & $0.63^{+0.09}_{-0.11}$  & $0.59^{+0.17}_{-0.17}$ & -- & -- & $0.70^{+0.11}_{-0.10}$ & $0.69^{+0.15}_{-0.22}$\\
NGC\,4321 & -- & -- & $0.45^{+0.10}_{-0.08}$  & $0.59^{+0.20}_{-0.16}$ & -- & -- & $0.55^{+0.09}_{-0.08}$ & $0.72^{+0.17}_{-0.18}$\\
NGC\,4536 & $0.84^{+0.36}_{-0.41}$ & -- & --  & $0.59^{+0.51}_{-0.32}$ & $0.85^{+0.23}_{-0.34}$ & -- & -- & $0.65^{+0.46}_{-0.36}$\\
NGC\,4579 & $0.35^{+0.13}_{-0.15}$ & -- & --  & $0.45^{+0.17}_{-0.19}$ & $0.38^{+0.21}_{-0.11}$ & -- & -- & $0.54^{+0.24}_{-0.23}$\\
\enddata
\tablecomments{The table lists the median and 16$^{\rm th}$ to 84$^{\rm th}$ percentile range.}
\end{deluxetable*}

\subsection{Comparison to other surveys}
%\begin{itemize}
%    \item How do trends relate to EMPIRE $R_{21}$?
%    \item How do trends relate to \citet{Leroy2022} $R_{21}$?
%\end{itemize}
%\subsection{Impact of flux calibration uncertainty}
Numerous mapping surveys that combine CO(1-0) and CO(2-1) at kpc-scales to investigate $\rtwo$ across nearby galaxies exist \citep{Leroy2013, denBrok2021, Yajima2021, Leroy2022}. An overview of the various galaxies and the particular mapping surveys that covered them can be found in \autoref{tb:mapping_surveys}. However, so far, these mapping surveys are based on single-dish observations, mostly using the either the IRAM 30m or the Nobeyama 45m telescope. Aggregated over the whole sample and combining the single-dish surveys, a consistent galaxy-wide $\rtwo$ average of around $0.65$ is found.  In \autoref{fig:comp_samples}, we compare the galaxy-integrated line ratio measurements of this sample to the ones presented in \citet{denBrok2021} (who find a CO(1-0) intensity weighted average of $\langle R_{21}\rangle{=}0.64\pm0.09$) and \citet{Leroy2022} (who report $\langle R_{21}\rangle{=}0.65_{-0.15}^{+0.18}$). In particular, the sample of galaxies covered in \citet{Leroy2022}, which spans a comparable range in total SFR and stellar mass to the survey in this study, exhibits a large galaxy-to-galaxy scatter (of around 0.3\, dex). We find that the average ratios across our galaxy sample are consistent -- within the scatter -- with the median value found by \citet{Keenan2024b}, who conducted unresolved, single-pointing, single-dish observations. Their study, which spans 122 galaxies, reports a median ratio of $\langle R_{21}\rangle{=}0.71_{-0.20}^{+0.25}$. The slightly higher ratio of 0.71 compared to our value of 0.64 may be attributed to the fact that their single-pointing observations are biased toward the brighter central regions (as the pointings are centered on the galaxy centers and might not include the full outskirts of the galaxy depending on its apparent size).

%\edit1
{For reference, we also the reported average line ratio values derived from GMC-resolved observations \citep{Sakamoto1994} in \autoref{fig:comp_samples}. For the Orion A cloud, a value of 0.66 is reported, while Orion B, on average, measures a ratio of 0.71. Within the cloud, the value of \rtwo\ is observed to increase to above unity towards H\textsc{ii} and OB association regions. While Orion A \& B are likely not representative of the entire extragalactic GMC population, these values still indicate that the average line ratios we observe within galaxies are mostly dominated by the intermediate and perioheral regions of GMCs.}

We %\edit1
{also present} a more detailed comparison when combining samples of single-dish observations in \autoref{fig:comp_samples}. The %\edit1
{central part of the figure} illustrates the line-of-sight $R_{21}$ value distribution across the sample of galaxies using different combinations of CO(1-0) and CO(2-1) maps (see \autoref{tb:mapping_surveys}). We notice that in particular when using the Nobeyama 45m CO(1-0) observations to derive $R_{21}$, the galaxy-wide average seems to be systematically lower by ${\sim}$0.1 dex. Based on this panel and the observed offset alone, it is not clear if this is a selection effect, since the $R_{21}$ distributions are not necessarily drawn from the same galaxies or even same field-of-view, as different surveys mapped different galaxies. In \autoref{fig:comp_literature_r21}, we additionally compare the relation of $R_{21}$ as function of the SFR surface using all sightlines from the sample observed in this study (presented in red) with the single-dish-based result presented in \citet{Leroy2022} (presented in black) and the correlation reported by \citet{Keenan2024a} for galaxy-integrated measurements (blue line). Two things are evident: (i) the scatter when only using ALMA CO(1-0) and CO(2-1) data is lower (despite a similar dynamical range of the galaxies' SFR and stellar mass in jibthe sample), but (ii) the relation of $R_{21}$ and SFR surface density shows a similar slope of around 0.1, despite the difference in the scatter for all three studies.

%This comparison underscores the SFR surface density constitutes a robust predictor for $R_{21}$ variation, closely aligning with earlier studies. This convergence suggest that $R_{21}$ values can now be more confidently predicted across diverse galactic environments for the range of SFR surface densities from 10$^{-3}$ to 10$^{0}$ M$_\odot\,\rm yr^{-1}\,kpc^{-2}$.

This comparison underscores that SFR surface density serves as a predictor for $R_{21}$ variation, in agreement with previous studies. While there is some scatter, approximately 0.5 dex, around the correlation for a given SFR, this relationship remains the most robust method currently available for predicting $R_{21}$ across diverse galactic environments, especially within the SFR surface density range\footnote{%\edit1
{We have employed a single power law given our limited dynamical range. We expect the trend to flatten at both high $\Sigma_{\rm SFR}$ (towards unity at $\Sigma_{\rm SFR}{\gtrsim}0.3$M$_\odot\,\rm yr^{-1}\,kpc^{-2}$) and low $\Sigma_{\rm SFR}$ (the ratio will likely not go below ${\sim}$0.3). Therefore, it is feasible to establish more complex fitting relations (see, e.g., \citealt{Leroy2022} and \citealt{Keenan2024b}, who use a broken power law fit with a flat trend above and below some threshold values). However, in our case, we do not include such a fit given the limited dynamic range.}} of 10$^{-3}$ to 10$^{0}$ M$_\odot\,\rm yr^{-1}\,kpc^{-2}$. Investigating the factors contributing to this scatter offers a promising direction for future research.

\begin{figure*}
    \centering
    \includegraphics[width=0.85\textwidth]{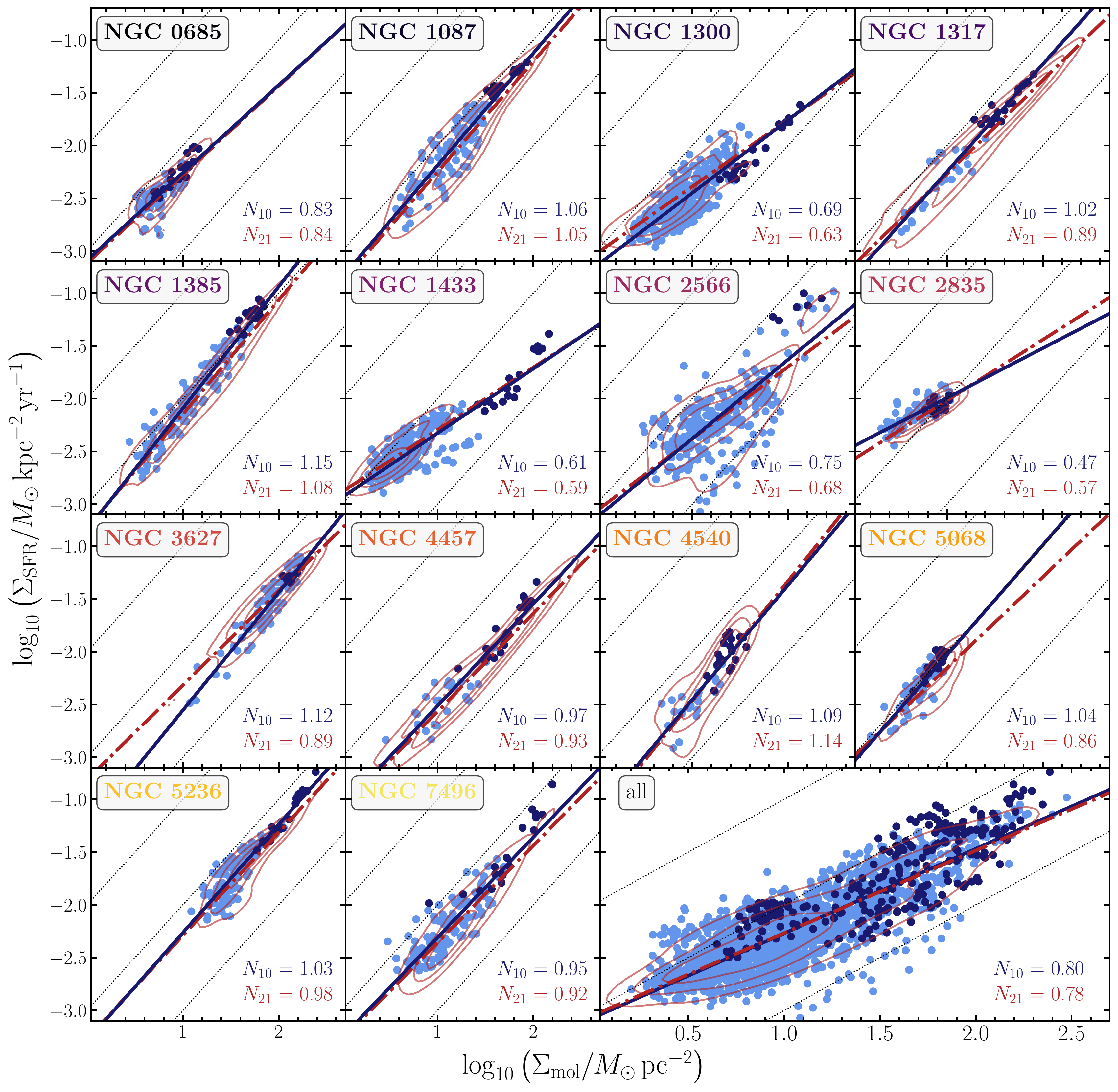}
    \caption{{\bf The Kennicutt-Schmidt relation}. The blue points show the correlation of the molecular gas surface density based on the CO(1-0) line intensity with SFR surface density. The dark blue points indicate central ($r{<}2\,\rm kpc$) sightlines. We fit a linear regression in log-log space and report the slope, which corresponds to the KS index, in each panel ($N_{10}$). The red contours show the distribution of the same sigthlines but using the CO(2-1) and a fixed $R_{21}{=}0.64$ to determine the molecular gas surface density. The red, dashed line shows the linear regression to the CO(2-1)-based points. The corresponding KS index, $N_{21}$, is indicated in each panel as well. The diagonal dotted lines indicate constant depletion time of 0.1, 1, and 10 Gyr from top to bottom. }
    \label{fig:ks_implication}
\end{figure*}

\begin{figure*}
    \centering
    \includegraphics[width=0.95\textwidth]{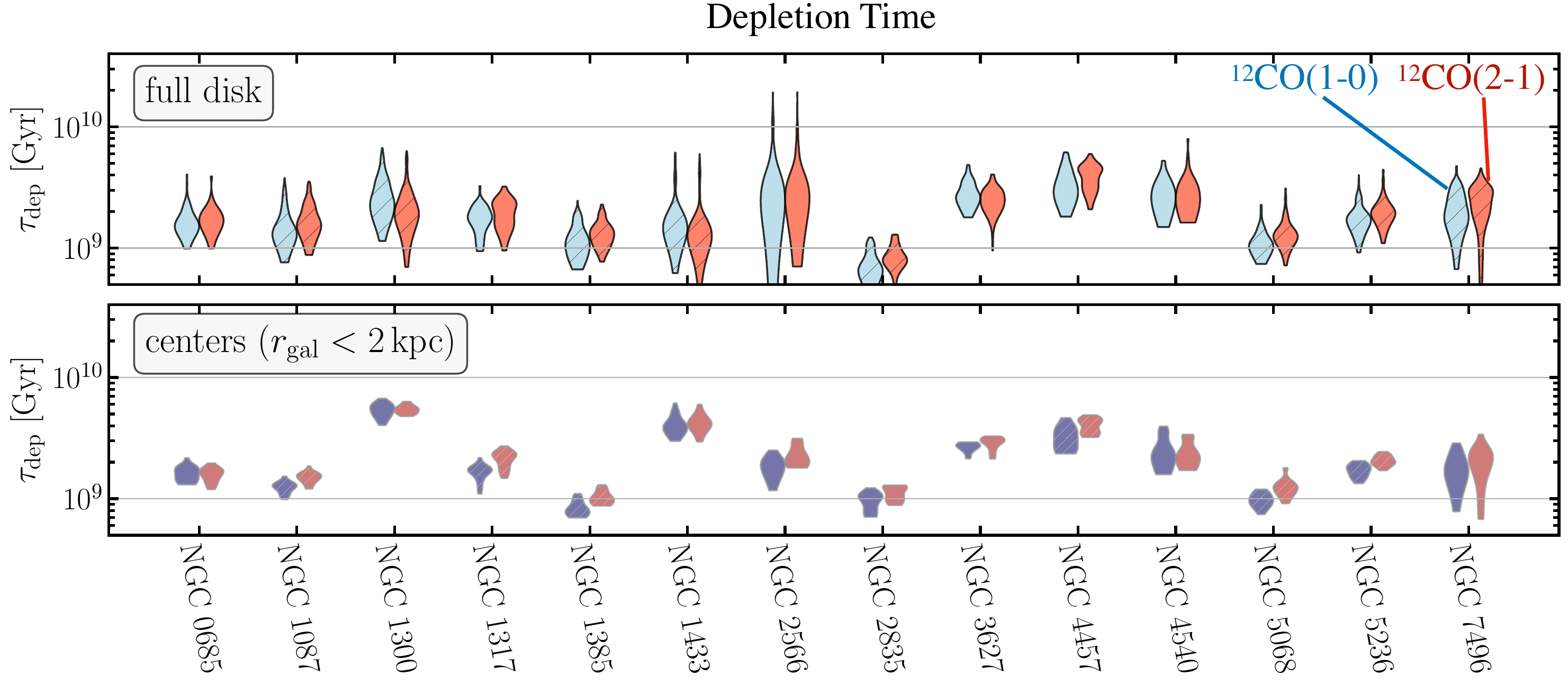}
    \caption{{\bf Galaxy-to-galaxy variation in depletion time}. These panels contrast the bulk molecular gas depletion time using either the CO(1-0), in blue, or CO(2-1) and a fixed $R_{21}$, in red. The top panel presents the distribution across the full galaxy disks. The bottom panel illustrates the distribution for the central galaxy points ($r_{\rm gal}{<}2$\,kpc). Where the distribution of depletion times when either using CO(1-0) or CO(2-1) differs significantly based on a KS test, they are presented using hatched violins.}
    \label{fig:depl_time}
\end{figure*}

%%%%%%%%%%%%%%%%%%%%%%%%%%%%%%%%%%%%%%%%%%%%%%%%%%%%%%%%%%%%%%%%%%%%%
%
%.     Discussion 
%
%%%%%%%%%%%%%%%%%%%%%%%%%%%%%%%%%%%%%%%%%%%%%%%%%%%%%%%%%%%%%%%%%%%%%
\section{Discussion } \label{sec:disc}

\subsection{Impact on resolved molecular gas scaling relations}

%\edit1
{
There is broad interest in the rate at which stars form from molecular gas in galaxies, and how this rate varies as a function of host galaxy and local environment \citep[e.g.,][]{Young1991,Young1996,Kennicutt2012}. One point motivating the study of \rtwo\ has been to understand how the CO line used for a particular study impacts results on this topic.}

%\edit1
{There are several formalisms adopted to study variations in SFR per unit molecular mass. This quantity is often expressed as the star formation efficiency of H$_2$ \citep[e.g.,][]{Young1996,Leroy2008} or the molecular gas depletion time \citep[e.g.,][]{Saintonge2022}, }
\begin{equation}
\label{eq:tdep}
\tau_{\rm dep}^{\rm mol} = \frac{\Sigma_{\rm mol}}{\Sigma_{\rm SFR}}
\end{equation}
%\edit1
{which captures the time that it would take star formation to consume the available molecular gas reservoir. One of the most popular framings has been to assume that the molecular gas surface density represents a controlling parameter and study the Kennicutt-Schmidt relation \citep[][]{Schmidt1959, Kennicutt1998} in its molecular form,} \begin{equation}
    \frac{\Sigma_{\rm SFR}}{M_\odot\,\rm yr^{-1}\,kpc^{-2}} = c\times \left( \frac{\Sigma_{\rm mol}}{M_\odot\,\rm pc^{-2}} \right)^N~.
\end{equation}
%\edit1
{Use of the molecular form of the Kennicutt-Schmidt relation reflects that molecular gas is the phase observed to correlate most directly with tracers of star formation at $\approx$kpc scales \citep[][]{wong2002,bigiel2008,Schruba2011}. The parameter $c$ describes an amplitude related to the star formation efficiency and gas depletion time, and the variation of star formation rate per unit gas as a function of $\Sigma_{\rm mol}$ is captured by the power law index, $N$ (hereafter refereed to as the KS index). Though focusing exclusively on $N$ misses the dependence of the star formation rate per molecular gas on any environmental factor other than $\Sigma_{\rm mol}$, $N$ has become a popular, if imperfect, shorthand for the influence of environment on the star formation process.}

%\edit1
{In \autoref{fig:ks_implication}, we compare the Kennicutt-Schmidt relation for the two transitions and report differences in the best-fit KS index $N$ for the two transitions. In these plots, we use a radially varying CO-to-H$_2$ conversion factor based on the prescription from \citet{Bolatto2013} and computed for the galaxies in our sample by \citet{Sun2022}. However, these value do not account for variations in the line ratio \citep[e.g.,][where this term is considered]{Schinnerer2024}.  We find KS indices ranging from $N=0.5$ to $1.12$, though we note that the adopted fitting technique, as well how upper limits are treated, impacts the precise KS index derivation \citep[e.g.,][]{Leroy2013}.To compare results for the two transitions, we fit the relation to the same points when using the CO(2-1) measurements and a fixed $R_{21}$. {We note that the derived KS indices should only be compared cautiously with literature results, as the measurements are sensitive to the specific methodology used for calculating the slopes.} }

%\edit1
{\autoref{fig:ks_implication} shows that for most galaxies, the choice of line has a modest impact on the estimated $N$. When combining all sightlines across all galaxies, the KS index only changes by $-2\%$, from $N_{10}{=}0.80$ to $N_{21}{=}0.78$. The impact on individual galaxies is larger but still usually modest. For 10 out of 14 targets, $N$ varies by less than $0.1$ or $\pm 10\%$ between CO(1-0) to CO(2-1). For 7 of these targets, the change in the KS index is less than 5\%. The other four galaxies that show more significant variations are NGC\,1317, NGC\,2835, NGC\,3627, and NGC\.5068. For 11/14 galaxies, the KS index decreases when using CO(2-1) instead of CO(1-0). This is expected, as we find elevated $\rtwo$ values at larger SFR surface densities (see \autoref{fig:r21_sfr_trends}). As a result, the assumption of a fixed line ratio will lead to overestimating the CO(2-1) derived molecular gas mass surface densities at larger SFR surface densities, which will flatten the KS relation and leading to lower KS index. The sense of our results is consistent with similar discussions in \citet{Yajima2021} and \citet{Leroy2022}, though we find smaller overall impact of line choice on $N$. See \citet{Sun2023} for a recent study of the the impact of a variety different SFR and $\alpha_{\rm CO}$ prescriptions on $N$ in the PHANGS--ALMA data set, also at kpc-scale resolution. } 

%\edit1
{In \autoref{fig:depl_time}, we show the effect of line choice on the estimated $\tau_{\rm dep}^{\rm mol}$ (\autoref{eq:tdep}) for whole galaxies and for only galaxy centers. Formally, we find that for 8 of 14 galaxies, the distribution of depletion times across the whole disk  differ significantly between the two lines when assuming a fixed \rtwo. However, the figure also shows that the variations of $\tau_{\rm dep}^{\rm mol}$ from galaxy to galaxy and from galaxy center to galaxy centers are mostly not a result of line choice. For the most part, these appear similar regardless of line choice. That is, a galaxy seen to have high $\tau_{\rm dep}^{\rm mol}$ using CO~(1-0) also shows high $\tau_{\rm dep}^{\rm mol}$ using CO~(2-1).}

%\edit1
{Figures \ref{fig:ks_implication} and \ref{fig:depl_time} therefore suggest that the impact of $\rtwo$ variation on the derived scaling relations are minimal for kpc-scale observations of main sequence galaxies. For individual galaxies, the bias induced by assuming using CO(2-1) and a fixed $\rtwo$ does generally not exceed 10\%. This is smaller than uncertainties in the CO-to-H$_2$ conversion factor, especially for galaxy centers \citep{Sandstrom2013,denBrok2023,Teng2023}. Despite their modest magnitude, as we have seen above physical variations in the line ratio do exist and should be accounted for to achieve best estimates of $N$ and $\tau_{\rm dep}^{\rm mol}$.}

%We therefore suggest to use an average line ratio of $\langle \rtwo\rangle{=}0.66$ for nearby star-forming spiral galaxies for kpc-scale CO(2-1) observations. For central sighlines, a larger line ratio of $\langle \rtwo\rangle^{\rm center}{=}0.75$ will capture the main but minor bias induced by the expected $\rtwo$ variation.

\subsection{How to predict $\rtwo$ {at kpc-scale}?}

Overall, assuming a fixed $\rtwo=0.64$ -- for either the galaxy-wide average or for kpc-sclae sightlines -- will not significantly bias the resulting molecular gas scaling relations for massive main sequence galaxies if dealing with relations across a sample.  For an improvement of accuracy, we suggest the following options:

(i) For unresolved galaxy observations, stellar mass serves as an effective predictor of variations in $\rtwo$ (see left panel of \autoref{fig:r21_galToGal_trends}). Our findings indicate that higher stellar mass galaxies tend to have lower $\rtwo$ ratios. Therefore, if integrated stellar mass measurements are available, the galaxy-wide average ratio can be approximated using the following empirical relation (valid for the stellar mass range of ${\sim}10^9{-}10^{11}\,M_\odot$):
\begin{equation}
    \log_{10}\langle R_{21}\rangle_{\rm gal} \approx -0.07{\pm}0.02 \log_{10}\left(\frac{M_*^{\rm gal}}{M_\odot}\right) + 0.57{\pm}0.09
\end{equation} 
Note that in particular for single-dish observations, the scatter is expected to be in the range of 0.1--0.2 dex due to additional flux calibration uncertainties. {Therefore, we recommend using the fixed line ratio of $\rtwo=0.64$ if flux calibration uncertainty is expected to be large (${>}10\%)$. }

(ii) In the context of resolved, kpc-scale observations, the star formation rate surface density ($\Sigma_{\text{SFR}}$) provides a robust predictor for $\rtwo$. Specifically, the relation indicates that regions with higher $\Sigma_{\text{SFR}}$ tend to exhibit higher $\rtwo$ ratios. Our study (see \autoref{table:fitting}) suggest the following relationship for SFR surface densities with $\rtwo$ {(valid for the SFR surface density range of ${\sim}10^{-3}{-}10^{-1}\,\rm M_\odot\,yr^{-1}\,kpc^{-2}$)}:
\begin{equation}
\footnotesize
    \log_{10} R_{21}^{\rm 1.7\,kpc} \approx 0.12^{+0.03}_{-0.03} \log_{10}\left(\frac{\Sigma_{\rm SFR}}{M_\odot\rm \, yr^{-1}\,kpc^{-2}}\right) + 0.06^{+0.07}_{-0.09}
\end{equation} 
%Note that a scatter of 0.1\,dex can be expected particularly if different galactic environments, such as center and disk are included. 
Note that a scatter of 0.1 dex can be expected, particularly when different galactic environments, such as the center and disk, are included.  Future work will investigate whether combining both predictors -- stellar mass and SFR surface density -- might improve the accuracy for both scales in predicting $\rtwo$.
%%%%%%%%%%%%%%%%%%%%%%%%%%%%%%%%%%%%%%%%%%%%%%%%%%%%%%%%%%%%%%%%%%%%%
%
%.     Conclusion
%
%%%%%%%%%%%%%%%%%%%%%%%%%%%%%%%%%%%%%%%%%%%%%%%%%%%%%%%%%%%%%%%%%%%%%
\section{Conclusions} \label{sec:conc}

Here we present new ALMA Morita ACA (7m + TP) CO(1-0) maps of 12 nearby galaxies ($D_{\rm gal}{\lesssim}$20\,Mpc). We combine this data set with two archival CO(1-0) maps and CO(2-1) ALMA data from the PHANGS-ALMA survey. We convolve all observations to a common physical resolution of 1.7\,kpc, which corresponds the the smallest common scale. With the combined CO(1-0) and CO(2-1) data, we investigate $\rtwo$ variation across and within the galaxies. In particular, we address the following science questions:

\begin{enumerate}
     \item \textit{To what {degrees do varying physical conditions in the gas, and uncertainties in the relative flux calibration contribute to previously observed galaxy-to-galaxy scatter in \rtwo}?}

      We find an overall galaxy-to-galaxy scatter of 0.06~dex in the global average $\rtwo$, which is lower than the ${>}$0.1~dex reported by previous studies with comparable sample sizes. We conclude that previous single-dish surveys particularly suffered from flux calibration uncertainties of up to 20\%, which significantly affected the measured galaxy-to-galaxy scatter. 
      
     \item \textit{On what magnitude does the CO line ratio $\rtwo$ vary at 1.7\,kpc physical resolution across the distinct environments of center, disk, arm, and interarm?}
     
     We find a 0.1~dex internal variation in $\rtwo$ within galaxies, which dominates over the 0.06~dex galaxy-to-galaxy scatter.  At a physical resolution of 1.7\,kpc, we observe an increase of $\rtwo$ towards the center of the galaxies, where the mean value increases to 0.75 (15\% increase). 
     
     \item \textit{How well can we parameterize the $\rtwo$ variation (e.g., as a function of SFR surface density) across and within nearby galaxies?}
     
     We investigate correlations of $\rtwo$ with galactocentric radius, SFR surface density, specific SFR, and metallicity. We find significant correlations, as quantified by the Kendall's $\tau$ correlation coefficient, with all four parameters. The most significant correlation is found with the SFR surface density. In logarithmic space, we find a linear regression with a slope of 0.12 when fitting to the binned, and 0.11 when fitting to all lines of sight. We note, however, that individual galaxies can show steeper or shallower correlations between the SFR surface density and $\rtwo$.
 \end{enumerate}

Overall, we find that using CO(2-1) and assuming a fixed $\rtwo$ does not significantly bias molecular gas mass estimates at kpc-scale resolution across large samples of galaxies. In contrast, systematic uncertainties in the flux calibration (particularly from single-dish telescope) and CO-to-H$_2$ conversion factor are more significant and can drive galaxy-to-galaxy scatter in molecular gas mass estimates as well as scatter within individual galaxies. We recommend an average  $\rtwo=0.64$, which is consistent with previous findings.
We note, however, that when only looking into an individual galaxy, the induced bias by a fixed line ratio can be important, especially in regions close to the galaxy center. Expanding the sample beyond the massive star-forming galaxy population along with obtaining GMC-scale resolved observations will be fundamental in obtaining further constraints and insights into the environmental dependent variation of $R_{21}$ in the ISM.
%%%%%%%%%%%%%%%%%%%%%%%%%%%%%%%%%%%%%%%%%%%%%%%%%%%%%%%%%%%%%%%%%%%%%
%
%.    Rest
%
%%%%%%%%%%%%%%%%%%%%%%%%%%%%%%%%%%%%%%%%%%%%%%%%%%%%%%%%%%%%%%%%%%%%%

%% Also note that the akcnowlodgment environment does not support long amounts of text. If you have a lot of people and institutions to acknowledge, do not use this command. Instead, create a new \section{Acknowledgments}.
\section*{Acknowledgments}
This work was carried out as part of the PHANGS collaboration.
JdB and EWK acknowledge support from the Smithsonian Institution as a Submillimeter Array (SMA) Fellow. JS acknowledges support by the National Aeronautics and Space Administration (NASA) through the NASA Hubble Fellowship grant HST-HF2-51544 awarded by the Space Telescope Science Institute (STScI), which is operated by the Association of Universities for Research in Astronomy, Inc., under contract NAS~5-26555. HAP acknowledges support from the National Science and Technology Council of Taiwan under grant 113-2112-M-032-014-MY3. A.K.L. gratefully acknowledge support from NSF AST AWD 2205628, JWST-GO-02107.009-A, and JWST-GO-03707.001-A and support by a Humbolt Research Award. FHL acknowledges support from the Scatcherd European Scholarship of the University of Oxford. AU acknowledges support from the Spanish grant PID2022-138560NB-I00, funded by MCIN/AEI/10.13039/501100011033/FEDER, EU.

This paper makes use of the following ALMA data: \\
\noindent ADS/JAO.ALMA\#2012.1.00650.S, \linebreak % (N628/M74)
ADS/JAO.ALMA\#2013.1.00803.S, \linebreak % (N5128/CenA)
ADS/JAO.ALMA\#2013.1.01161.S, \linebreak % (N1365 + N5236/M83)
ADS/JAO.ALMA\#2015.1.00121.S, \linebreak % (N5236/M83)
ADS/JAO.ALMA\#2015.1.00782.S, \linebreak % (N1313 + N7793)
ADS/JAO.ALMA\#2015.1.00925.S, \linebreak % (pilot low mass)
ADS/JAO.ALMA\#2015.1.00956.S, \linebreak % (pilot high mass)
ADS/JAO.ALMA\#2016.1.00386.S, \linebreak % (N5236/M83)
ADS/JAO.ALMA\#2017.1.00392.S, \linebreak % (low mass follow-up)
ADS/JAO.ALMA\#2017.1.00766.S, \linebreak % (early-type)
ADS/JAO.ALMA\#2017.1.00886.L, \linebreak % (large program)
ADS/JAO.ALMA\#2018.1.01321.S, \linebreak % (N253, N300, Circinus)
ADS/JAO.ALMA\#2018.1.01651.S, \linebreak % (main sample follow-up)
ADS/JAO.ALMA\#2018.A.00062.S, \linebreak % (ACA-only nearby)
ADS/JAO.ALMA\#2019.1.01235.S, \linebreak % (local sample follow up)
ADS/JAO.ALMA\#2019.2.00129.S, \linebreak % (N1068)
ADS/JAO.ALMA\#2022.1.01479.S, \linebreak % (Jakob CO 1-0)
ADS/JAO.ALMA\#2015.1.01538.S,
\linebreak % (NGC 3627 CO 1-0)
ADS/JAO.ALMA\#2017.1.00079.S,
\linebreak % (NGC 5236 CO 1-0)

ALMA is a partnership of ESO (representing its member states), NSF (USA), and NINS (Japan), together with NRC (Canada), NSC and ASIAA (Taiwan), and KASI (Republic of Korea), in cooperation with the Republic of Chile. The Joint ALMA Observatory is operated by ESO, AUI/NRAO, and NAOJ. The National Radio Astronomy Observatory is a facility of the National Science Foundation operated under cooperative agreement by Associated Universities, Inc.

\vspace{5mm}
\facilities{Atacama Large Millimeter/submillimeter Array (\textit{ALMA})}

\software{astropy \citep{2013A&A...558A..33A,2018AJ....156..123A, Astropy_v5},  
numpy \citep{harris2020array},  PyStructure \citep{PyStructure_v3}, %\edit1
{PHANGS-ALMA pipeline \citep[v3.1; ][]{Leroy2021}}
          }

\appendix
\section{ALMA Flux recovery}
\label{app:flux_recov}
\begin{figure}
    \centering
    \includegraphics[width=0.55\textwidth]{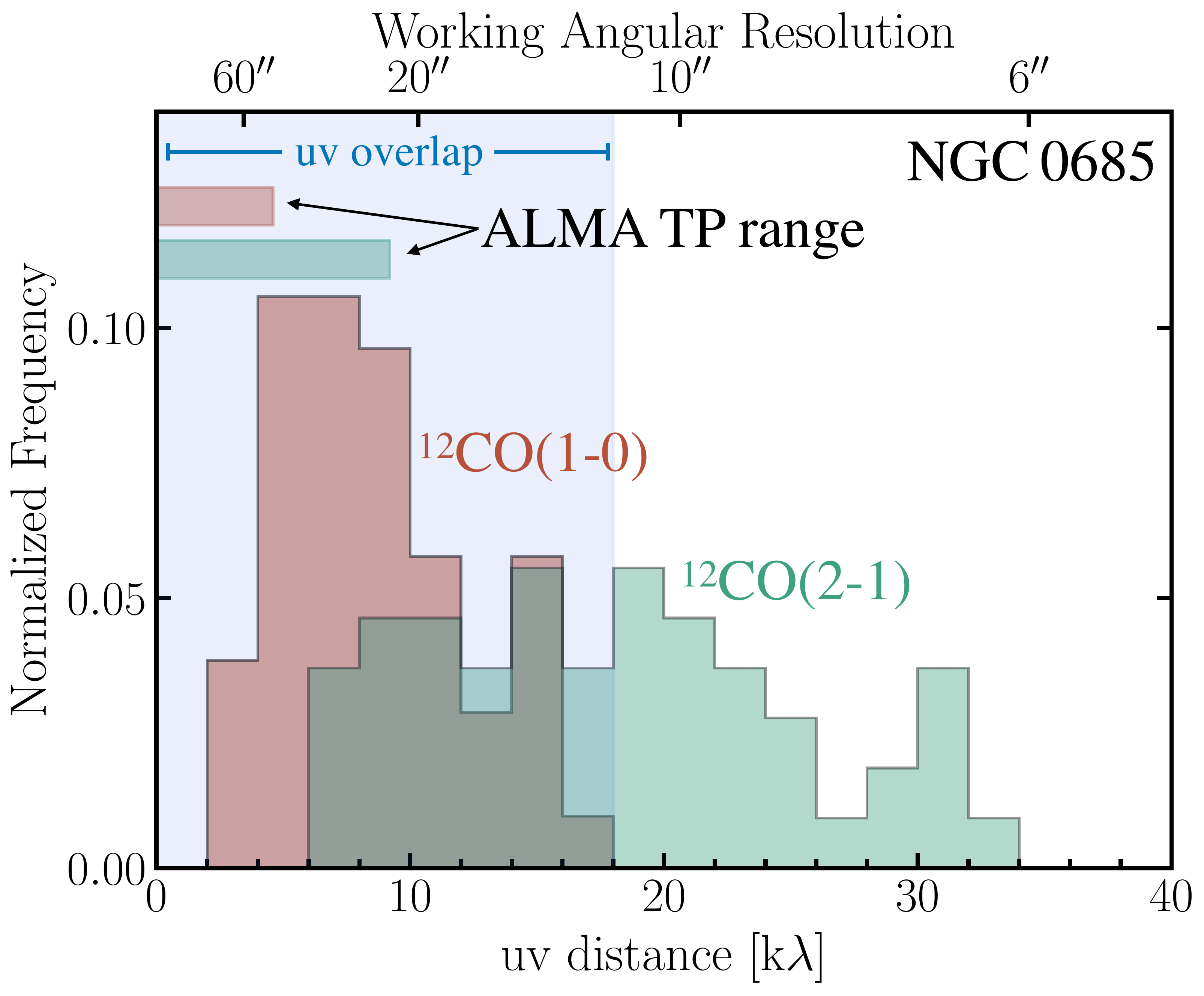}
    \caption{\textbf{Overlap in $uv$-distance for CO(1-0) and CO(2-1) 7m+TP observations.} The normalized histograms illustrate the distribution of the channel and time averaged amplitudes of the visibilities for the ALMA 7m observations of CO(1-0) in brown and CO(2-1) in green. In addition, the range of the TP observations is indicated by the bars. The blue shaded regions represents the overlap in terms of $uv$-distance out to ${\sim}18$k$\lambda$ (${\sim}14''$).   }
    \label{fig:uv_overlap}
\end{figure}

{A main challenge with interferometric ALMA data remains the reliability of the combined 7m+TP observations and their ability to recover CO emission, accounting for the $u-v$ coverage and calibration uncertainties. We note, however, that significant prior work has been devoted to precisely quantify the issue of flux recovery in the context of nearby galaxy CO emission. \citet{Leroy2021} have investigated the calibration and reliability of ALMA TP observations by comparing these fluxes of the four separate TP antennas, finding rms variations of 2–4\% due to flux calibration uncertainties. Furthermore, \citet{Leroy2021} and \citet{Neumann2023_stacking} demonstrated that combined 7m+TP observations reliably recover ${>}$90\% of the total flux in simulated datasets for nearby face-on galaxies. In \autoref{fig:uv_overlap}, we plot the distribution of the ALMA 7m visibilities (averaged over all channels and time) in terms of $uv$-distance for CO(1-0) and CO(2-1) for one galaxy in our sample (here NGC\,0685).  When including the ALMA TP observations, the overlap for CO(1-0) and CO(2-1) extends to ${\sim}18$\,k$\lambda$ in $uv$-distance, corresponding to an angular resolution of ${\sim}14''$.}

\section{Comparing different $R_{21}$ Fitting Prescriptions}
\label{app:fitting_presc}

\begin{figure}
    \centering
    \includegraphics[width=0.55\textwidth]{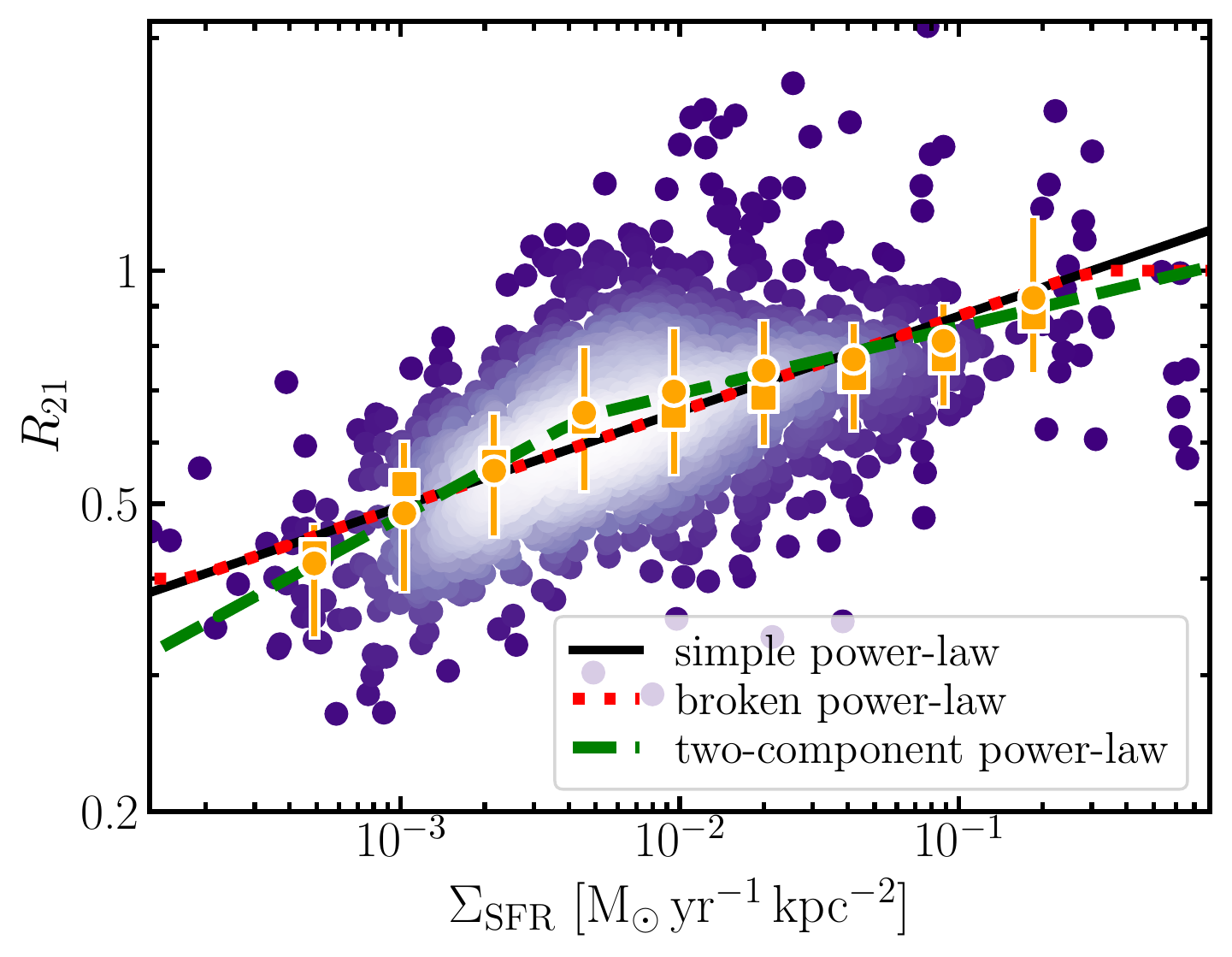}
    \caption{\textbf{Contrasting different $R_{21}$ fitting prescriptions.} In the paper, we employ a simple power law (black line) to describe the scaling relation between the line ratio and the SFR surface density. Alternate functional forms of the scaling relations have been suggested, capturing different, physically motivated aspects. For instance, \citet{Leroy2022} propose a broken power law (red-dotted line), which describe the scaling relation using a power law with a maximum and minimum line ratio. In addition, a two-component power law can also be used, which describes the contribution from a star-forming ans non-star-forming H$_2$ gas component. We list the respective reduced $\chi^2$ values for these functional forms in \autoref{tab:sample}. }
    \label{fig:sfr_prescriptions}
\end{figure}

\begin{deluxetable}{clc}

 \tablecaption{ Different functional forms of the scaling relation of the line ratio ($y\equiv\log_{10}(R_{21})$) and the SFR surface density ($x\equiv\log_{10}(\Sigma_{\rm SFR})$). The reduced $\chi^2$ is derived from the fit shown in \autoref{fig:sfr_prescriptions}. \label{tab:sample}}

 \tablehead{
 \colhead{Prescription} & \colhead{Functional Form} & \colhead{Reduced $\chi^2$} 
 }

 \startdata 
 Simple Power law & $y=m\cdot x + q$ & 0.077\\
 Broken Power law & $y = \begin{cases}y_{\rm low} & x<x_{\rm low}\\ y_{\rm low}+m\cdot(x-x_{\rm low}) & x_{\rm low}\le x \le {x_{\rm high}}\\ 0.0 & x>x_{\rm high}\end{cases}$ & 0.090 \\
 Two-component Power law & $y=\begin{cases}
     m_1\cdot x + q & x<x_0\\
     m_2\cdot x + (m_1 - m_2)\cdot x_0 & x\ge x_0 
 \end{cases}$& 0.075
 \enddata

 \vspace{-0.5cm}
 %\tablecomments{sample comment}

\end{deluxetable}

{In Section \ref{sec:res_meas}, we use a power law function to fit the relation between the SFR surface density ($\Sigma_{\rm SFR}$) and the line ratio (see \autoref{fig:r21_relation_all}). However, the underlying relation is likely more complex than a simple, single exponential law, and hence alternative functional forms have been proposed in previous studies. Critically, however, we are limited with a narrow range of SFR surface density, over which $R_{21}$ scales following an exponential law. For completeness, we provide here a comparison of such alternative prescriptions (see \autoref{tab:sample} for overview). We note however, that given the limited dynamical range in SFR surface density, such parametrization are not clearly justified by the data as the statistical improvement (in terms of reduced $\chi^2$) is insignificant. \autoref{fig:sfr_prescriptions} illustrates the different prescriptions.}

{An alternative, physically motivated, scaling relation for the line ratio is proposed by \citet{Leroy2022} (hereafter called \emph{broken power law}). This functional form adopts a maximum line ratio at 1 and a minimal line ratio at $y_{\rm low}$. The motivation is that we do not expect the line ratio to exceed, on kpc-scales, significantly above unity, and we do not expect the line ratio to decrease arbitrarily low, even at low SFR surface densities. In between, the ratios are described with an exponential power law. When fitting such a broken power law to our data, we find a slightly higher reduced $\chi^2$ of 0.090 (compared to 0.077 using the simple power law). Finally, we can also describe the $R_{21}$ relation using a two-component power law approach. The two-component approach is motivated by separating the gas into a SF-related vs. non-SF related H$_2$ phase. Calculating the reduced $\chi^2$ value of 0.075, we find that it is only insignificantly better than the simple power law. Therefore, given in particular the limitation by the dynamical range the data cover in terms of SFR surface density, the simple power law prescription is used. }

\bibliography{references}{}
\bibliographystyle{aasjournal}


%% Include this liene if you are using the \added, \replaced, \deleted

\end{document}